# Engineering crystal structure and spin-phonon coupling in $Ba_{1-x}Sr_xMnO_3$

Bommareddy Poojitha, Ankit Kumar, Anjali Rathore, Devesh Negi, and Surajit Saha*

Department of Physics, Indian Institute of Science Education and Research, Bhopal 462066, India

* Corresponding author. Email address: surajit@iiserb.ac.in (Surajit Saha)

**Abstract**

The interplay between different degrees of freedom such as charge, spin, orbital, and lattice has received a great deal of interest due to its potential to engineer materials properties and their functionalities for device applications. In this work, we have explored the crystallographic phase diagram of $Ba_{1-x}Sr_xMnO_3$ and studied the correlation between two degrees of freedom, namely phonons and spins using magnetization and inelastic light scattering measurements. The system undergoes a series of crystallographic phase transitions 2H ➔ 9R ➔ 4H as a function of doping (Sr) as observed by X-ray diffraction measurements. Investigation of their temperature-dependent magnetization reveals a para- to antiferro-magnetic transition for all the compositions. An $E_g$ phonon in the 9R phase and an $E_{1g}$ phonon in the 4H phase involving Mn/O-vibrations, show anomalous temperature-dependence in the antiferromagnetic phase arising due to spin-phonon coupling.

**Introduction**

Antiferromagnetic (AFM) materials offer several advantages over conventional ferromagnets (FM) due to multiple reasons, such as, AFMs are more common than FMs, ordering is permitted in each magnetic symmetry group, allows different electronic ground states (AFMs can be found as metals, semiconductors, insulators, semimetals, or superconductors whereas FMs are mostly metallic), and so on [1-3]. The ultrafast dynamics and unique properties of AFM materials make them superior in applications, giving birth to new fields of research such as AFM spintronics, AFM opto-spintronics, and AFM piezo-spintronics [1, 4-8]. On the other hand, a strong coupling between the magnetic degree of freedom and phonons in AFM materials gives an opportunity to tune the magnetic ground state and trigger the ground state into multiferroic [9, 10]. In this context, divalent perovskite manganites ($AMnO_3$: A = Ca, Sr, and Ba) are extensively studied due to their multifunctional properties arising from the presence of different order parameters with strong coupling among them [11, 12]. Especially, BaMnO$_3$ attracts attention due to its polymorphic nature [13, 14]. It can be stabilized in a variety of structural phases (2H, 4H, 6H, 8H, 10H, 9R, 12R, 15R, 21R, 27R, and 33R, where the integer and letters (H and R) represent the number of layers and crystal symmetry (hexagonal and rhombohedral) respectively) by tuning parameters such as synthesis conditions and strain (by chemical doping in bulk and external strain in substrate supported thin films). The magnetic properties of different polymorphs of BaMnO$_3$ have been extensively studied [15]. They exhibit para- to antiferro-magnetic transition with $T_N$ ~ 59-270 K. On the other hand, SrMnO$_3$ is an AFM insulator which can be stabilized either in cubic or 4H phase, where the hexagonal (4H) phase has a higher Neel temperature ($T_N$ ~ 280 K) [16] than its cubic counterpart ($T_N$ ~ 230 K) [11]. Wang et al. reported first-principles calculations on SrMnO$_3$/BaMnO$_3$ superlattices revealing the tunability of its electric and magnetic ground states using epitaxial strain [17]. The multiferroic ground state above 100 K with giant magnetoelectric coupling was theoretically predicted for strained AMnO$_3$ thin films, owing to spin-phonon coupling (SPC) [10, 11, 18-19], but, the requirement of high strain (3 - 5%) makes it difficult to achieve in practice (in thin films/superlattices). The alternative way to apply strain is chemical doping where a partial substitution of $Sr^{+2}$ ($r$ = 1.44 Å) in place of $Ba^{+2}$ ($r$ = 1.61 Å) introduces compressive strain on the lattice that leads to strain-induced ferroelectricity (FE). FE in single crystals of Ba$_{1-x}$Sr$_x$MnO$_3$ (0.55 ≤ x ≤ 0.50) below ~ 410 K was observed by Sakai et al. [20, 21]. Therefore, in Ba$_{1-x}$Sr$_x$MnO$_3$ when both the FE and magnetic orders coexist, presence of spin-phonon interactions can induce magnetoelectric coupling [10, 11, 18-19].

Here, we have extensively studied the correlation of phonons with magnetic degrees of freedom in polymorphs of Ba$_{1-x}$Sr$_x$MnO$_3$: x = 0.00 to 0.60. The powder X-ray diffraction (PXRD) results reveal a series of structural phase transitions 2H ➔ 9R ➔ 4H as a function of Sr concentration. Furthermore, a structural phase transition from high temperature *P6$_3$mc* symmetry to the low temperature (below 130 K) *P6$_3$cm* symmetry is observed for x = 0.00 and 0.05 from temperature-dependent PXRD results. However, we did not observe any signatures of change in the crystal symmetry for the compositions with x ≥ 0.10 with temperature (in the range of 90 - 400 K). Our Raman measurements show that the $E_g$ (ω ~ 344 cm$^{-1}$) phonon related to Mn displacements exhibit anomalous behaviour with temperature in the 9R phase compositions (x = 0.10 to 0.30). On the other hand, the $E_{1g}$ mode at ~ 420 cm$^{-1}$ in 4H phase exists only at low temperatures, showing anomalous temperature-dependence for x = 0.40 and 0.50, while exhibiting normal temperature-dependance for x = 0.60. The observed phonon anomalies in $Ba_{1-x}Sr_xMnO_3$: x = 0.10 to 0.50 are attributed to variable SPC. The presence of spin-phonon interactions was reported by us earlier for one of these compositions $Sr_{1-x}Ba_xMnO_3$ (x = 0.9) [22]. The strength of the SPC ($\lambda$) has been estimated using Lockwood's method [23, 24].

**Experimental details**

Polycrystalline samples of $Ba_{1-x}Sr_xMnO_3$: x = 0.00, 0.05, 0.10, 0.125, 0.15, 0.20, 0.30, 0.40, 0.50, and 0.60 have been synthesised by solid-state reaction route. High purity powders (Sigma-Aldrich) of $BaCO_3$, $SrCO_3$, and $MnO_2$ were used as precursors. A well-mixed powder in stoichiometric ratio was ground and then calcined at 800°C, 900°C, and 1000°C for 24 hours each time and sintered at 1200°C for 12 hours with intermediate grindings. PXRD measurements were performed in PANalytical Empyrean X-ray diffractometer attached with Cu K$_\alpha$ radiation of wavelength 1.5406 Å and liquid nitrogen-based Anton Paar TTK 450 heating stage. Chemical compositions were determined using energy dispersive X-ray spectroscopy (EDAX) technique equipped with high-resolution field emission scanning electron microscope (HR-FESEM) (Zeiss ULTRA Plus). Raman spectroscopic measurements were performed using LabRAM HR Evolution (Horiba Scientific) Raman spectrometer equipped with 532 nm laser source (Nd:YAG), 50× long working distance microscope objective, and Peltier cooled charge-coupled device (CCD) detector. Raman spectra were collected in the temperature range of 80 - 400 K using the Linkam stage (model HFS600E-PB4). DC magnetization measurements were done using Quantum Design SQUID-VSM (Superconducting Quantum Interference Device attached with Vibrating Sample Magnetometer).

**Results and discussion**

Structural and magnetic properties:

The powder X-ray diffraction (PXRD) patterns on polycrystalline powders of $Ba_{1-x}Sr_xMnO_3$: x = 0.00 to 0.60 were collected at room temperature and analyzed by using Rietveld refinement method in High Score Plus (see Fig. S1 in supplementary material). The crystallographic phase diagram including the corresponding unit cells (drawn in VESTA (Visualization of Electronic and STructural Analysis) [25]), are shown in Fig. 1. The refinement results reveal that the compounds with x = 0.00 and 0.05 stabilize in 2H phase with the space group *P6$_3$mc* (No. 186). The crystal structure of 2H-BaMnO$_3$ can be viewed as columns of face-shared MnO$_6$ octahedra interleaved with chains of $Ba^{+2}$ ions, both running parallel to the c-axis [26-28]. The $Mn^{+4}$ ion occupies the corner positions of the unit cell and is surrounded by six $O^{-2}$ ions forming MnO$_6$ octahedron and $Ba^{+2}$ ions sit at the centre. Compositions with x = 0.10 to 0.15 exhibit 9R-type hexagonal structure (rhombohedral axis) with *R-3m* (No. 166) of symmetry. The unit cell of 9R phase has a stacking sequence (chh)$_3$ where c and h represent cubic and hexagonal layers [26-27, 29]. In other words, it consists of Mn$_3$O$_{12}$ units (formed by three face-shared MnO$_6$ octahedra) connected to each other at their corners. On the other hand, the compositions with x = 0.20, 0.30, and 0.40 stabilize in mixed-phase compounds containing the 9R and 4H unit cells. The weight fractions of both phases are extracted from the double-phase refinement method using High Score Plus software, as shown in Table S1 in Supplementary material. The single-phase 4H structure is observed for compounds with higher Sr-content with x = 0.50 and 0.60. The 4H crystal structure belongs to the hexagonal family with the crystal symmetry *P6$_3$/mmc* (Space group No. 194). It has a stacking sequence of (ch)$_2$ and consists of dimers of face-shared octahedra linked through their corners (see inset of Fig. 1) [26-27, 30]. The Rietveld refined lattice parameters of respective crystal phases and refinement residuals for all compositions are listed in Table S1 (plotted in Fig. S2 in supplementary material), which are consistent with earlier reports [22, 31-32] (bond lengths and bond angles are shown in Fig. S3). The decreasing trend in lattice parameters is seen as a function of doping (Sr$^{+2}$) in all crystallographic phases which is in support to the fact that the partial substitution of smaller Sr$^{+2}$ions ($r$ = 1.44 Å) in place of Ba$^{+2}$ ($r$ = 1.61 Å) leads unit cells to shrink [33]. The chemical compositions of as-grown samples were verified using EDAX spectroscopic measurements. The obtained values match well with the expected compositions within the instrumental resolution. No impurity elements or additional phases in the sample (see Table S2, Figs. S4 and S5 in supplementary material) are detected.

Figure 2 shows the magnetization as a function of temperature (MT) for Ba$_{1-x}$Sr$_x$MnO$_3$: x = 0.00 to 0.60 in the temperature range T = 100 to 400 K (data down to 2 K are shown in Fig. S6 in supplementary material). MT curves were measured in both zero-field cooling (ZFC) and field cooling (FC) cycles by applying 500 Oe magnetic field. The MT data of x = 0.00 (2H-BaMnO$_3$) show no signature of long-range AFM ordering in the temperature range of 80 - 400 K. However, it shows a broad feature in the MT curve below ~ 220 K, indicating the presence of short-range magnetic interactions at lower temperatures. Cussen *et. al.,* [34] reported an antiferromagnetic ordering below 59 K for the 2H phase (see Supplementary material for details) from neutron powder diffraction measurements. The compositions with x ≥ 0.05 exhibit a *cusp*-like feature in MT, indicating a transition from high-temperature paramagnetic to low-temperature AFM phase. The Neel temperature (T$_N$: 231 - 266 K) for each composition is determined from the temperature derivative curves of magnetization ($\frac{dM(T)}{dT}$ vs $T$: shown in Fig. S7 in supplementary material) and listed in Table I. The observed AFM transition arises from the 180° Mn-O-Mn superexchange interactions between $Mn^{+4}$ ions sitting in the corner-shared octahedra. The non-linear superexchange interactions between $Mn^{+4}$ ions located in the face-shared octahedra result a relatively weak or negligible ferromagnetic (FM) interactions due to minimal overlap between the 3d orbitals of $Mn^{+4}$ and $O^{-2}$ 2p orbitals (Fig. S3 in supplementary material).

Raman active-phonons in $Ba_{1-x}Sr_xMnO_3$:

According to factor group analysis [35], the 2H-$BaMnO_3$ with *P6₃mc* symmetry (space group No. 186) leads to 14 Raman active phonons at the Γ-point with the irreducible representations $Γ_{2H} = 4A_1 + 5E_1 + 5E_2$. The 9R-phase has crystal symmetry *R-3m* (space group No. 166) for which the Raman active phonon modes at the centre of the Brillouin zone predicted by group theory are $Γ_{9R} = 4A_{1g} + 5E_g$. Similarly, the crystal symmetry for 4H phase is *P6₃/mmc* (space group No. 194) with the irreducible representations for the Γ-point Raman phonons: $Γ_{4H} = 2A_{1g} + 2E_{1g} + 4E_{2g}$. The Raman spectra of $Ba_{1-x}Sr_xMnO_3$: x = 0.00 to 0.60 collected at room temperature are shown in Fig. 4. The spectra match well with the reported data [22, 31-32], exhibiting all characteristic modes expected for the respective crystal phases. The phonon parameters (frequency ω, linewidth Γ, and intensity A) are obtained by analyzing the spectra using the Lorentzian function. The symmetries of phonon modes in all the compositions are assigned based on the SYMMODES-Bilbao Crystallographic Server [35] and previous reports [22, 31-32] (see Tables II-IV in main text and S3 in supplementary material). As discussed above, the compositions with x = 0.20 to 0.40 exhibit mixed-phase containing both the 9R and 4H unit cells. Since Raman spectroscopy is a local probe, we could choose to collect spectra corresponding to the major crystal phase (9R spectra for x = 0.20 & 0.30 and 4H spectra for x = 0.40) in the mixed-phase compounds. To understand the thermal evolution of the phonons in the respective phase and probe the possible correlations with the magnetic properties, we have collected Raman spectra at various temperatures across the magnetic transitions.

The phonon frequencies of (2H) $Ba_{0.95}Sr_{0.05}MnO_3$ are blue-shifted as compared to those of (2H) $BaMnO_3$ (see Figs. S8 and S9 in supplementary material) which can be attributed to the compression of the unit cell due to smaller ionic radius of the dopant ($Sr^{2+}$) with respect to $Ba^{2+}$, as also observed from PXRD results (Table S1 in supplementary material). The ω (Γ) decreases (increases) with increasing temperature for all modes, as expected due to phonon anharmonicity. Figures S10-S14 show Raman spectra of 9R and 4H compositions (x: 0.10 to 0.60) collected at a few temperatures. Figures S15-S19 show phonon parameters (frequency and linewidth) as a function of temperature for x = 0.10 to 0.30 analyzed for the 9R phase, where phonons are labelled from P1-P8. Similarly, the phonon parameters for 4H phase compositions are shown in Figs. S20-S22, where phonons are labelled from Q1 - Q7. The ω (Γ) of most of the phonons decreases (increases) with increasing temperature in all (2H, 9R and 4H) the compositions. However, the frequency of the P3 phonon (related to Mn vibrations) in 9R compositions shows anomalous behaviour at low temperatures, i.e., the mode softens with decreasing temperature (Fig. 4(a)). The frequency of mode Q4 which involves Mn displacements also shows an anomalous behaviour with temperature for the compositions with x = 0.40 and 0.50, whereas it exhibits normal (anharmonic) behaviour for x = 0.60 (Fig. 4(a)). Notably, the Q4 appears as a new mode at low temperatures, well below $T_N$ (see Fig. S14 in supplementary material). These compositions don't show any change in the crystal symmetry in the temperature range T: 90-400 K, as evidenced from our temperature-dependent PXRD measurements (to be discussed later). Hence, structural phase transition can be ruled out as the possible origin of this new mode (Q4) at low temperatures. To be noted that the low-frequency phonons corresponding to the 4H phase have very low intensity at room temperature and are prominent only at low temperatures (below 200 K). To understand the origin of observed phonon anomalies, we have analyzed temperature-dependent results using Klemens anharmonic model described below.

In general, the temperature-dependent phonon frequency is given by [36-39]:

$$\omega(T) = \omega_{anh}(T) + \Delta\omega_{el-ph}(T) + \Delta\omega_{sp-ph}(T) \quad (1)$$

The first term $\omega_{anh}(T)$ is the phonon frequency due to anharmonicity. The term $\Delta\omega_{el-ph}(T)$ is due to electron-phonon coupling which is absent in the electrically insulating $Ba_{1-x}Sr_xMnO_3$ compounds studied here. $\Delta\omega_{sp-ph}(T)$ is the change in the phonon frequency due to SPC. The temperature-dependence of phonon frequency due to cubic anharmonicity is given by [38, 39]:

$$\omega_{anh}(T) = \omega_0 + A\left[1 + \frac{2}{\left(e^{\frac{\hbar\omega_0}{2k_BT}}-1\right)}\right] \quad (2)$$

Similarly, the change in linewidth due to the cubic anharmonicity is:

$$\Gamma_{anh}(T) = \Gamma_0 + C\left[1 + \frac{2}{\left(e^{\frac{\hbar\omega_0}{2k_B T}} - 1\right)}\right] \qquad (3)$$

where, $\omega_0$ and $\Gamma_0$ are frequency and linewidth of the phonon at 0 K, $A$ and $C$ are cubic anharmonic coefficients for $\omega$ and $\Gamma$, respectively, $\hbar$ is reduced Planck constant, $k_B$ is Boltzmann constant and $T$ is the variable temperature. See Figs. S12-S20 in supplementary material for details on the experimental data for $\omega$ and $\Gamma$. In addition to the anomalous temperature-dependence of frequency of P3 and Q4, the modes P4, P7, P8 (for x = 0.10 to 0.30) and Q7 (for x = 0.40 to 0.60) also show a clear finite deviation from the expected anharmonic trend (Eqns. 2 & 3). To verify the possible association of crystal structure in phonon anomalies, we have performed temperature-dependent PXRD measurements. The PXRD patterns at a few temperatures and temperature-dependent lattice parameters for all the compositions are given in supplementary material (Figs. S24-S27). Lattice parameters of x = 0.10 to 0.50 show a positive thermal expansion [40], maintaining the same crystal symmetry in the studied temperature range. Hence, the observed phonon anomalies (in P3, P4, P7, P8, Q4, and Q7 modes) below $T_N$ are attributed to the presence of SPC in the compositions with x = 0.10 to 0.50. On the other hand, appearance of a new mode (Q4) well below $T_N$ in 4H compositions (x: 0.40 to 0.60) is possibly due to the presence of a weak magnetostriction in 4H lattice [41]. To be noted that if magnetic degree of freedom is correlated with the elastic (lattice) degree of freedom, spontaneous strain may occur (locally) in the lattice at the magnetic transition temperature leading to (weak) magnetostriction. One possible reason for the appearance of a new mode (Q4) well below $T_N$ in 4H compositions (x = 0.40 to 0.60) can be the presence of a weak magnetostriction in the 4H lattice [41]. Since we could not see clear signatures of structural changes from PXRD measurements, the magnetostriction in these compositions may be too weak to be detectable in PXRD data and, therefore, neutron diffraction and/ striction measurements may be required to verify this possibility which are beyond the scope of this work.

To recall, the AFM order arises from the superexchange interactions between $Mn^{+4}$ ions [42]. When phonons, involving Mn/O atomic vibrations, interact with spins in a magnetic material, magnetic energy cause the renormalization in phonon parameters ($\omega$, $\Gamma$, A) below $T_N$ [36, 37]. Figure 4(b) (and Figs. S15-S22 in the supplementary material) shows the linewidth as a function of temperature. A clear deviation in Γ from the anharmonic behaviour (Eq. 3) below $T_N$, further corroborates the presence of SPC in $Ba_{1-x}Sr_xMnO_3$ with $x \geq 0.10$.

Spin-phonon coupling:

The temperature-dependence of phonon frequency ($\omega$), in the presence of spin-phonon interactions, below $T_N$ can be written as [36]:

$$\omega(T) = \omega_{anh}(T) + \lambda <S_i.S_j>$$

where, the first and second terms represent $\omega$ due to phonon anharmonicity (given in Eq. 2) and the contribution from the spin-phonon interactions. $\lambda$ is the coefficient which gives the strength of SPC. The deviation in $\omega$ (see Figs. 4(a, b) and Figs. S15-S22 in supplementary material) can be estimated using the relation $\Delta\omega_{sp-ph}(T) = \omega(T) - \omega_{anh}(T)$ (see Figure 5 for the mode P3 and Q4). The strength of SPC ($\lambda$) is estimated using the relation [23, 24]:

$$\Delta\omega_{sp-ph} = -\lambda <S_i.S_j> = -\lambda\,\Phi(T)S^2 \qquad (4)$$

where $\Phi$ is the short-range order parameter and $S$ is the spin. Lockwood *et. al.*, [23] theoretically estimated the $\Phi(T)$ for $S = 2$ ($FeF_2$) and $S = 5/2$ ($MnF_2$) AFM systems using mean-field and two-spin cluster approximations. Since the estimates of $\Phi$ is not very sensitive to the value of spin ($S$), we have used the $\Phi$ estimated by Lockwood *et. al.,* [36] for $BaMnO_3$ where $S = 3/2$ ($Mn^{4+}$) as it was also used for $S = 1$ (NiO and $NiF_2$) AFMs [24, 36]. Thus, the $\lambda$ is estimated using the relation [24]:

$$\lambda = -\frac{\omega(T_{Low}) - \omega_{anh}(T_{Low})}{[\Phi(T_{Low}) - \Phi(2T_N)]S^2} \qquad (5)$$

where $\omega(T_{Low})$ and $\omega_{anh}(T_{Low})$ are experimentally observed and anharmonic estimated value of phonon frequency, respectively, at the lowest temperature measured ($T_{Low} \sim 80$ K in our case). The obtained values for $\lambda$ using Eq. 5 are listed in Table V (and Table S4 in Supplementary material). The modes P3 (in $x = 0.10$ to $0.30$) and Q4 (in $x = 0.40$ and $0.50$) have a strong SPC ($\lambda \sim 1.0$ to $2.7$ cm$^{-1}$) which can be compared to values reported in a few known systems such as manganite (4H) $Sr_{0.6}Ba_{0.4}MnO_3$ ($\lambda \sim 2.2$ cm$^{-1}$[32]), $Sr_2CoO_4$ ($\lambda \sim 3.5$ cm$^{-1}$ [43]), NiO nanoparticle ($\lambda \sim 3.1$-$3.5$ cm$^{-1}$ [44]), $MnF_2$ ($\lambda \sim 0.4$ cm$^{-1}$[23]), $FeF_2$ ($\lambda \sim 1.3$ cm$^{-1}$[23]), $ZnCr_2O_4$ ($\lambda \sim 3.2$ to $6.2$ cm$^{-1}$[45]), $La_2CoMnO_6$ ($\lambda \sim 1.7$ to $2.1$ cm$^{-1}$ [46], $Pr_2CoMnO_6$ ($\lambda \sim 0.51$ to $1.61$ cm$^{-1}$ [46, 47]), and $Cr_2Ge_2Te_6$ ($\lambda \sim 0.24$ to $1.2$ cm$^{-1}$ [48]). In comparison, the modes P4, P7, P8, and Q7 exhibit a lower strength for the SPC.

## Conclusions

In summary, we have synthesized a series of polycrystalline $Ba_{1-x}Sr_xMnO_3$: x = 0.00 to 0.60 using solid-state reaction method. A few phonon modes associated with Mn/O displacements in the 9R and 4H phase compositions are identified to get renormalized below the AFM transition temperature ($T_N$) due to spin-phonon coupling. We could see a tunable spin-phonon coupling by varying the Sr-content in $BaMnO_3$ which is associated with the respective changes in the crystallographic phase. We believe that the $Ba_{1-x}Sr_xMnO_3$ can be a good candidate for technological applications in the field of AFM spintronics and/or multiferroics.

## Acknowledgement

Authors acknowledge CIF, IISER Bhopal for PXRD, SQUID-VSM, and EDAX research facilities, B. P. acknowledges the University Grant Commission (UGC) for fellowship, D. N acknowledges the CSIR for fellowship (09/1020(0139)/2018-EMR-I), and S. S. acknowledges DST/SERB (project Nos. ECR/2016/001376 and CRG/2019/002668) and Nano-mission (Project No. SR/NM/NS84/2016(C)) for research funding. Support from DST-FIST (Project No. SR/FST/PSI195/2014(C) is also thankfully acknowledged. Authors acknowledge Mr. Saheb Karak and Ms. Ankita Ram (Department of Physics, IISER Bhopal) for their help with the Raman and PXRD measurements.

## Credit authorship contribution statement:

**Bommareddy Poojitha:** Conceptualization, Data curation, Formal analysis, Investigation, Methodology, Project administration, Visualization, Writing - original draft. **Ankit Kumar, Anjali Rathore, and Devesh Negi:** Methodology, Data curation. **Surajit Saha:** Conceptualization, Funding acquisition, Resources, Software, Supervision, Validation, Visualization, Writing - review & editing.

Table I. Neel temperature ($T_N$) for each composition of $Ba_{1-x}Sr_xMnO_3$: x = 0.00 to 0.60.

| x | $T_N$ (K) | x | $T_N$ (K) |
|---|---|---|---|
| 0.00 | 59 K [Ref. 28] | 0.20 | 264 |
| 0.05 | 231 | 0.30 | 264 |
| 0.10 | 262 | 0.40 | 263 |
| 0.125 | 263 | 0.50 | 264 |
| 0.15 | 263 | 0.60 | 266 |

Table II. Symmetry assignment of the Raman active phonon modes and their frequencies in $cm^{-1}$ unit) in $Ba_{1-x}Sr_xMnO_3$: x = 0.00 & 0.05 (2H phase) at room temperature.

| $\omega$ (cm$^{-1}$) at 300 K | | Mode/ Symmetry | $\omega$ (cm$^{-1}$) at 300 K | | Mode/ Symmetry |
|---|---|---|---|---|---|
| $BaMnO_3$ | $Ba_{0.95}Sr_{0.05}MnO_3$ | | $BaMnO_3$ | $Ba_{0.95}Sr_{0.05}MnO_3$ | |
| 96 | 96 | M1 | 425 | 425 | M7 |
| 97 | 98 | M2 | 484 | 484 | M8 ($A_1$) |
| 118 | 118 | M3 ($E_2$) | 524 | 524 | M9 ($E_2$) |
| 405 | 403 | M4 | 634 | 637 | M10 ($A_1$) |
| 411 | 411 | M5 ($E_2$) | 657 | 657 | M11 ($E_1$) |
| 420 | 418 | M6 | | | |

Table III. Symmetry assignment of the Raman active phonon modes and their frequencies in cm$^{-1}$ unit) in Ba$_{1-x}$Sr$_x$MnO$_3$: x = 0.10 to 0.30 (9R phase) at room temperature.

| $\omega$ (cm$^{-1}$) in 9R-Ba$_{1-x}$Sr$_x$MnO$_3$ | | | | | Mode/ Symmetry | Atoms participate in vibration |
|---|---|---|---|---|---|---|
| x = 0.1 | x = 0.125 | x = 0.15 | x = 0.2 | x = 0.3 | | |
| 100 | 102 | 102 | 102 | 102 | P1 ($E_g$) | Ba/Sr |
| 255 | 256 | 256 | 256 | 256 | P2 ($A_{1g}$) | Ba/Sr |
| 344 | 346 | 346 | 347 | 347 | P3 ($E_g$) | Mn |
| 408 | 407 | 408 | 407 | 407 | P4 ($A_{1g}$) | Mn |
| 424 | 425 | 426 | 426 | 429 | P5 ($E_g$) | O |
| 531 | 531 | 532 | 534 | 537 | P6 ($E_g$) | O |
| 563 | 563 | 564 | 565 | 564 | P7 ($A_{1g}$) | O |
| 637 | 637 | 638 | 638 | 636 | P8 ($A_{1g}$) | O |
| 656 | 654 | 655 | 658 | 663 | P9 ($E_g$) | O |

Table IV. Symmetry assignment of the Raman active phonon modes and their frequencies in cm$^{-1}$ unit) in Ba$_{1-x}$Sr$_x$MnO$_3$: x = 0.40 to 0.60 (4H phase) at room temperature.

| ω (cm$^{-1}$) in 4H-Ba$_{1-x}$Sr$_x$MnO$_3$ | | | Mode/ Symmetry | Atoms participate in vibration |
|---|---|---|---|---|
| x = 0.4 | x = 0.5 | x = 0.6 | | |
| 112* | 114 | 115 | Q1 ($A_{1g}$) | Ba |
| 238* | 237 | 238 | Q2 ($E_{2g}$) | Mn |
| 329* | 332 | 333 | Q3 ($E_{1g}$) | Mn |
| 419* | 423* | 419* | Q4 ($E_{1g}$) | Octahedral tilting |
| 431 | 434 | 437 | Q5 ($E_{2g}$) | Octahedral Bending |
| 540 | 544 | 548 | Q6 ($E_{2g}$) | asymmetric octahedral stretching (O) |
| 618 | 623 | 619 | Q7 ($E_{2g}$) | O |
| 634 | 637 | 640 | Q8 ($A_{1g}$) | symmetric octahedral stretching (O) |

Note: Values marked with * are at 120 K

Table V. Spin-phonon coupling constant (λ) for a mode involving Mn-vibrations having $E_g$ symmetry (P3) in the 9R phase (x = 0.10 to 0.30) and $E_{1g}$ symmetry (Q4) in the 4H phase (x = 0.40 and 0.50) of Ba$_{1-x}$Sr$_x$MnO$_3$.

| x | 0.10 | 0.125 | 0.15 | 0.20 | 0.30 | 0.40 | 0.50 |
|---|---|---|---|---|---|---|---|
| λ (cm$^{-1}$) | 1.8 | 1.9 | 1.3 | 1.8 | 2.7 | 1.0 | 1.2 |

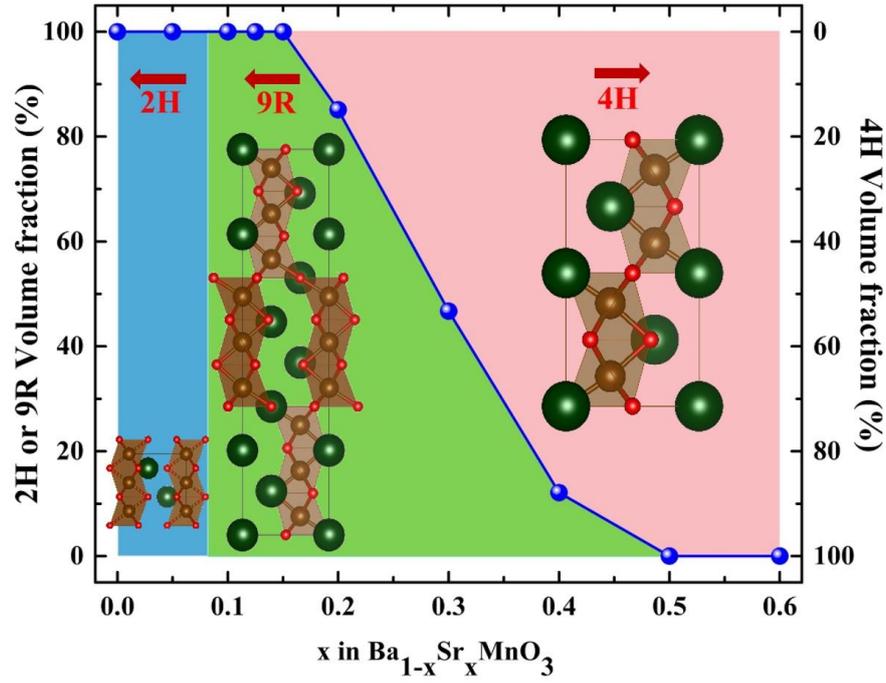

Figure 1. (Colour online) Crystallographic phase diagram of Ba$_{1-x}$Sr$_x$MnO$_3$: x = 0.00 to 0.60. Unit cells are shown in respective compositional range.

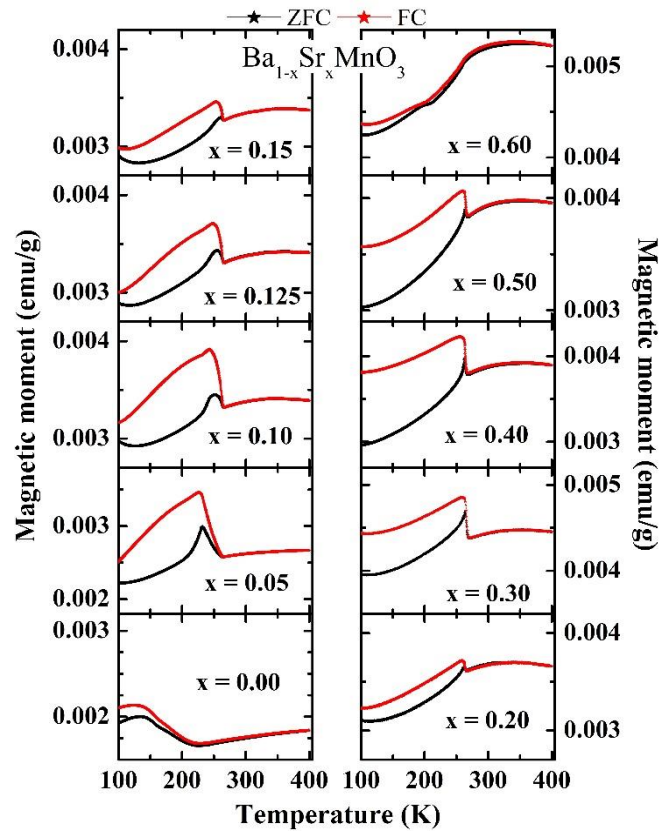

Figure 2. (Colour online) Magnetization as a function of temperature for Ba$_{1-x}$Sr$_x$MnO$_3$: x = 0.00 to 0.60.

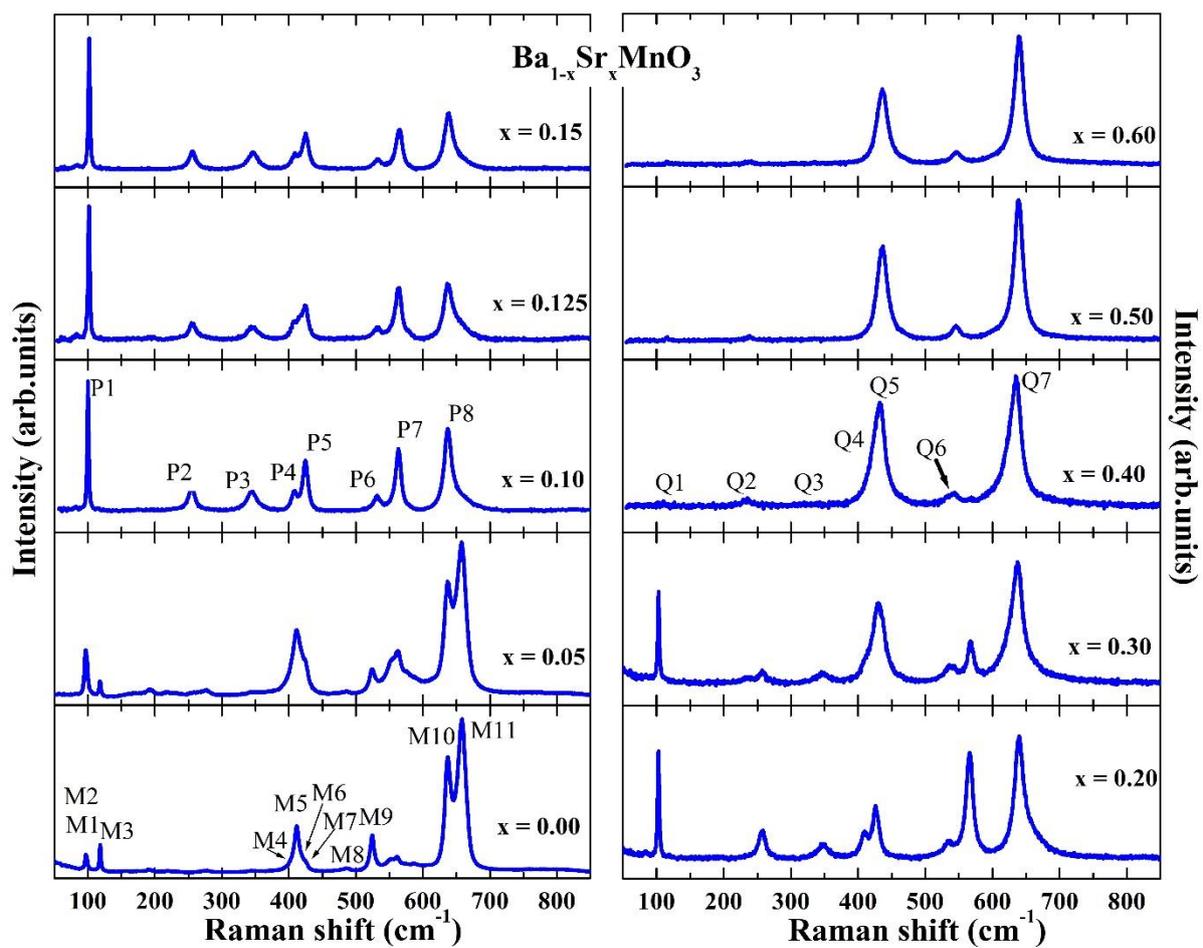

Figure 3. (Colour online) Raman spectra of $Ba_{1-x}Sr_xMnO_3$: x = 0.00 to 0.60 collected at room temperature.

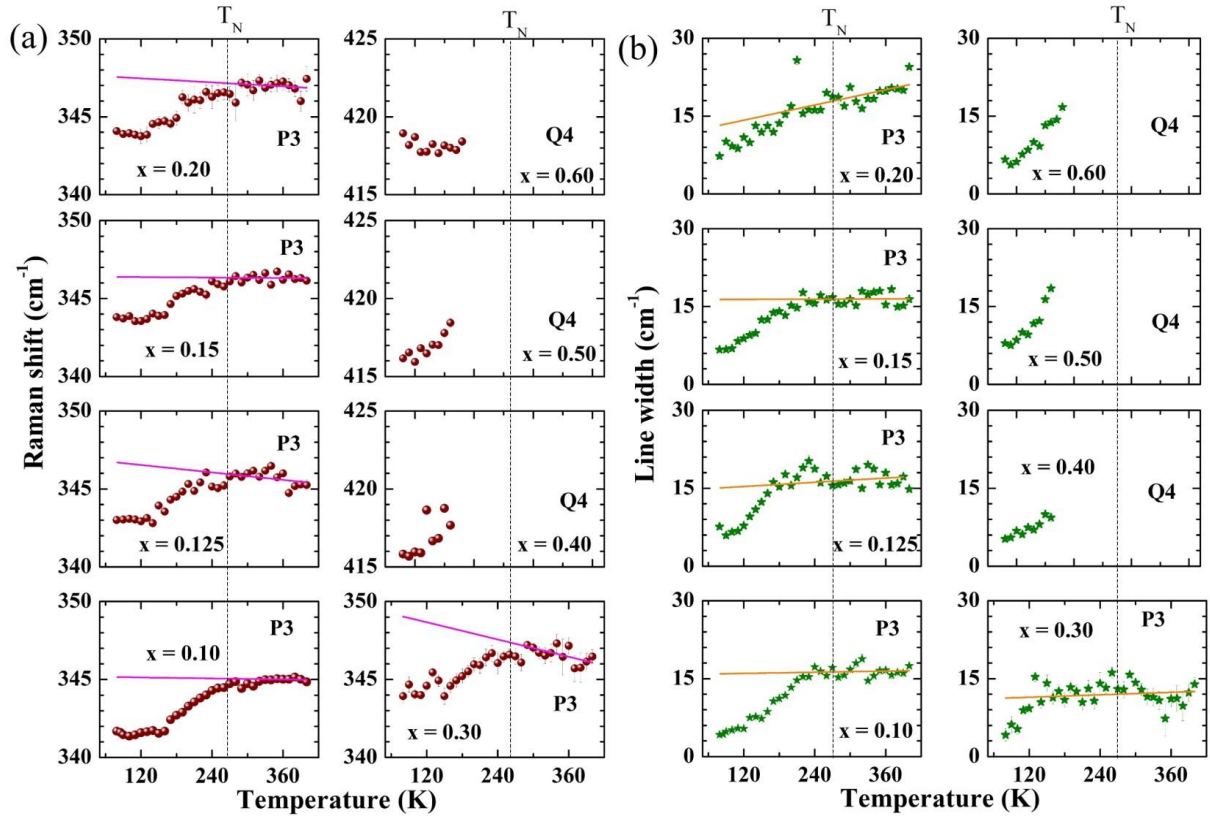

Figure 4. (Colour online) Frequency (a) and linewidth (b) of the phonons P3 and Q4 corresponding to the 9R and 4H phase, respectively. Solid lines (Magenta and orange) represent anharmonic fitting (Eqns. 2 and 3) for frequency and linewidth, respectively.

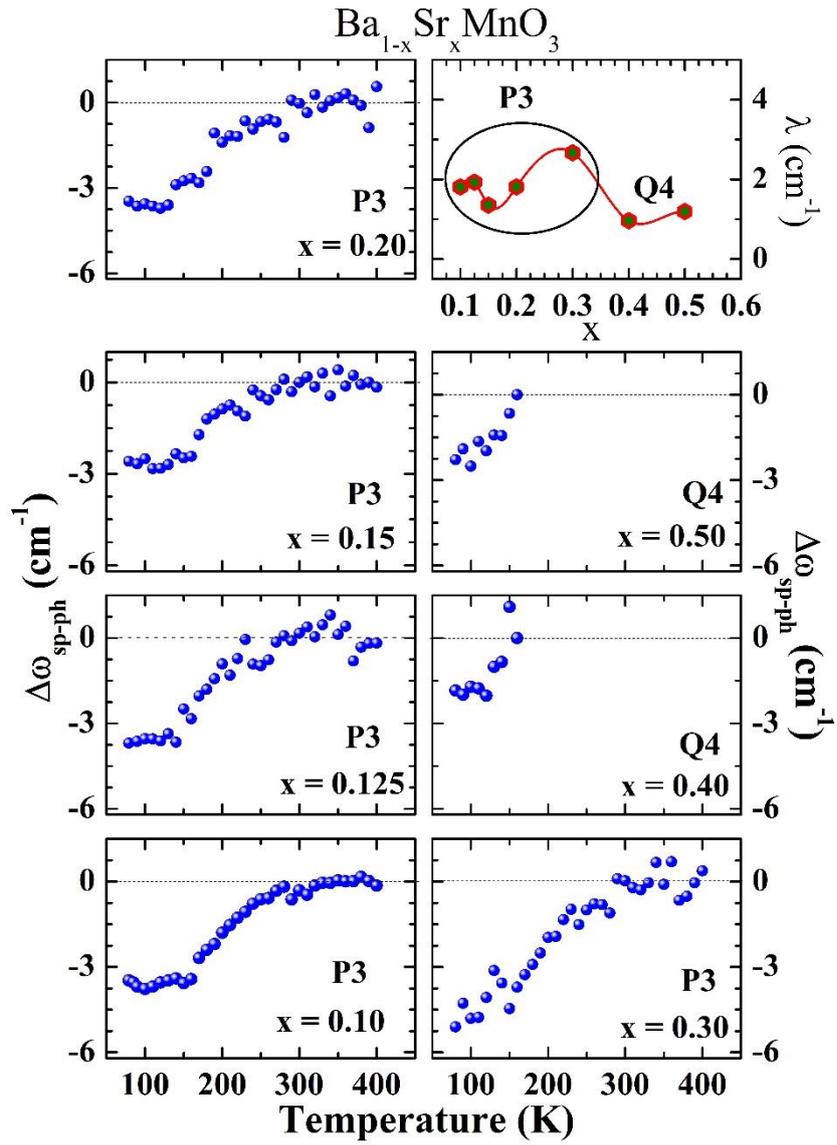

Figure 5. (Colour online) The variation of $\Delta\omega_{sp-ph}(T)$ with temperature in $Ba_{1-x}Sr_xMnO_3$: x = 0.10 to 0.50 and the spin-phonon coupling constant ($\lambda$) as a function of Sr content.

# Supplementary material

# Engineering crystal structure and spin-phonon coupling in $Ba_{1-x}Sr_xMnO_3$

Bommareddy Poojitha, Ankit Kumar, Anjali Rathore, Devesh Negi, and Surajit Saha*

Department of Physics, Indian Institute of Science Education and Research, Bhopal 462066, India

This supplementary material contains additional data on X-ray diffraction, magnetic, and Raman spectroscopy measurements of $Ba_{1-x}Sr_xMnO_3$. The figures and respective details are given below.

## Structural characterization: X-ray diffraction

Figure S1 shows X-ray diffraction patterns for each composition of $Ba_{1-x}Sr_xMnO_3$. The refinement residuals are given in Table S1.

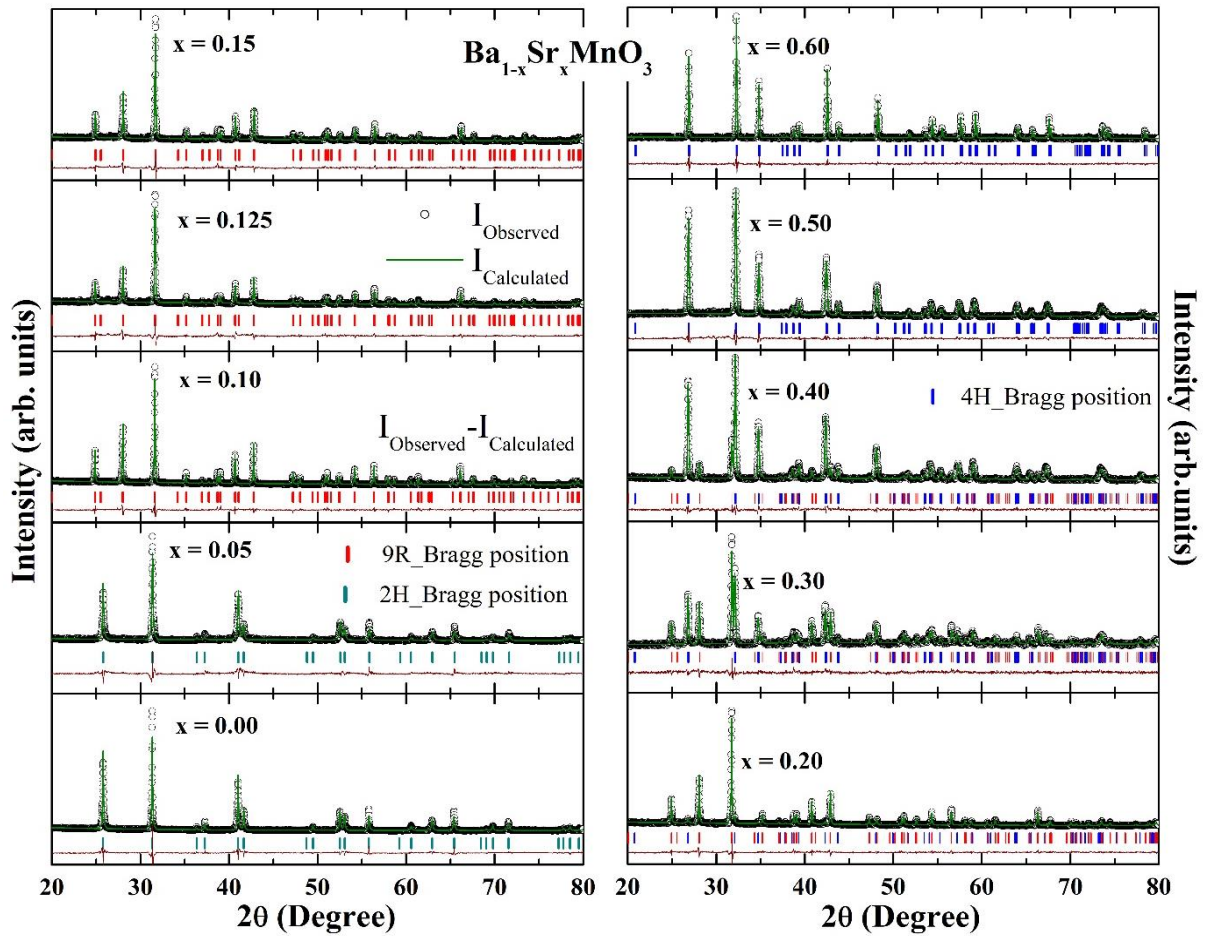

Figure S1. X-ray diffraction patterns along with Rietveld refinement for each composition of $Ba_{1-x}Sr_xMnO_3$.

Phase fractions

The weight fractions of the respective crystal phases at each composition are calculated from Rietveld refinement (Table S1) using High-score Plus software based on the equation,

$$w_p = (SZMV)_p / \sum_i (SZMV)_i \quad\quad\quad\quad\quad\quad (S1)$$

where, $w_p$ is weight fraction of phase $p$, S is the scale factor, Z is the number of formula units per unit cell, V is unit cell volume, and $i$ is index running over all phases.

Lattice parameters

Figure S2 shows the lattice parameters corresponding to 9R and 4H phases as a function of doping. Bond lengths of Mn-Mn across face-shared and corner-shared octahedra are plotted as a function of doping in Figure S3.

Table S1. Phase fractions and corresponding lattice parameters for each composition in $Ba_{1-x}Sr_xMnO_3$ extracted from X-ray diffraction data.

| Sample | Phase | Lattice parameters | | Refinement residuals | | |
|---|---|---|---|---|---|---|
| | | a (Å) | c (Å) | $R_p$ | $R_{wp}$ | GoF |
| x = 0 | 2H | 5.7000(8) | 4.8161(9) | 5.2636(4) | 7.5170(5) | 1.6502(4) |
| x = 0.05 | 2H | 5.6946(5) | 4.8191(4) | 5.7328(6) | 7.9595(2) | 1.7756(0) |
| x = 0.1 | 9R | 5.6445(0) | 20.897(0) | 3.7696(3) | 5.2071(3) | 0.9357(5) |
| x = 0.125 | 9R | 5.6409(3) | 20.886(9) | 4.4286(9) | 6.0574(7) | 1.2298(2) |
| x = 0.15 | 9R | 5.6362(2) | 20.887(3) | 3.8858(5) | 5.3527(9) | 1.2022(5) |
| x = 0.2 | 9R (85.1%) | 5.6309(5) | 20.886(7) | 3.0090(2) | 4.2050(2) | 0.8717(4) |
| | 4H (14.9%) | 5.5712(2) | 9.1529(9) | | | |
| x = 0.3 | 9R (46.7%) | 5.6280(8) | 20.882(3) | 2.2210(3) | 2.8456(5) | 1.5022(3) |
| | 4H (53.3%) | 5.5690(6) | 9.1521(7) | | | |
| x = 0.4 | 9R (12.1%) | 5.6280(7) | 20.882(7) | 1.7892(0) | 2.2874(5) | 1.5102(5) |
| | 4H (87.9%) | 5.5633(9) | 9.1495(3) | | | |
| x = 0.5 | 4H | 5.5506(9) | 9.1405(9) | 1.8444(8) | 2.4508(8) | 2.2960(1) |
| x = 0.6 | 4H | 5.5324(4) | 9.1317(4) | 1.6970(1) | 2.2843(8) | 1.3401(3) |

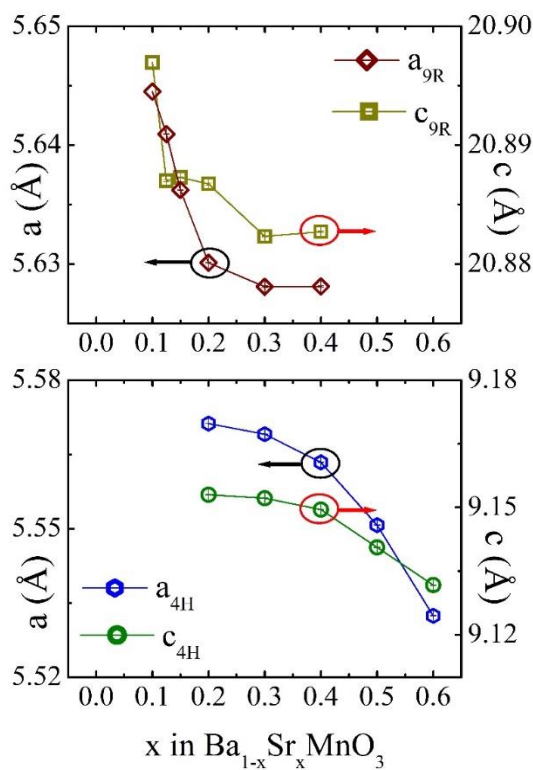

Figure S2. Lattice parameters as a function of Sr concentration in $Ba_{1-x}Sr_xMnO_3$. Error bars are within the symbol size.

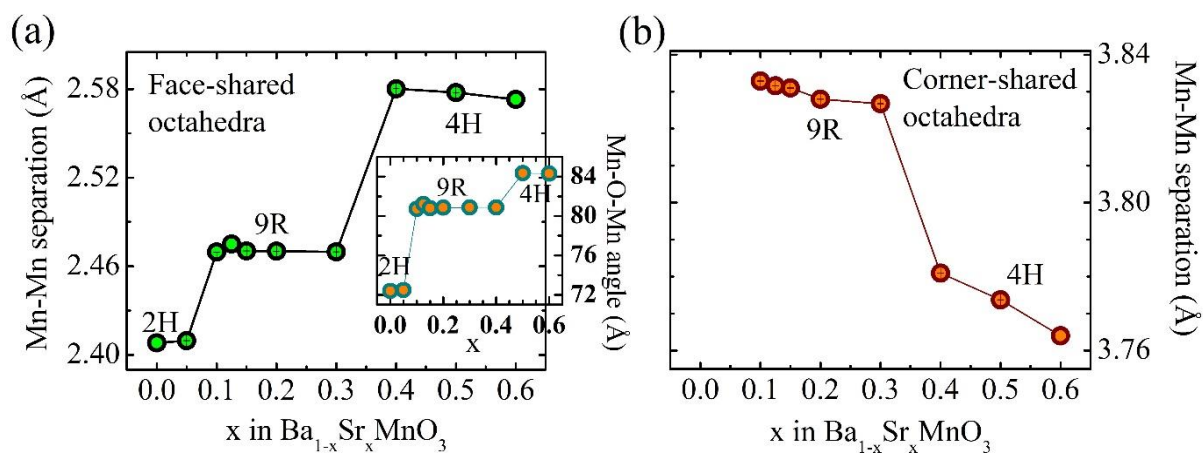

Figure S3. Mn-Mn separation across (a) face-shared and (b) corner-shared octahedra as a function of Sr concentration in $Ba_{1-x}Sr_xMnO_3$. Mn-O-Mn bond angle across face-shared octahedra is plotted in inset of (a) and is 180° across corner-shared octahedra.

**Chemical composition: EDAX measurements**

EDAX spectra (Figs. S4 and S5) confirm that the chemical compositions agree well with the expected stoichiometry. Obtained composition values are given in Table S2 which match reasonably with the expected x values.

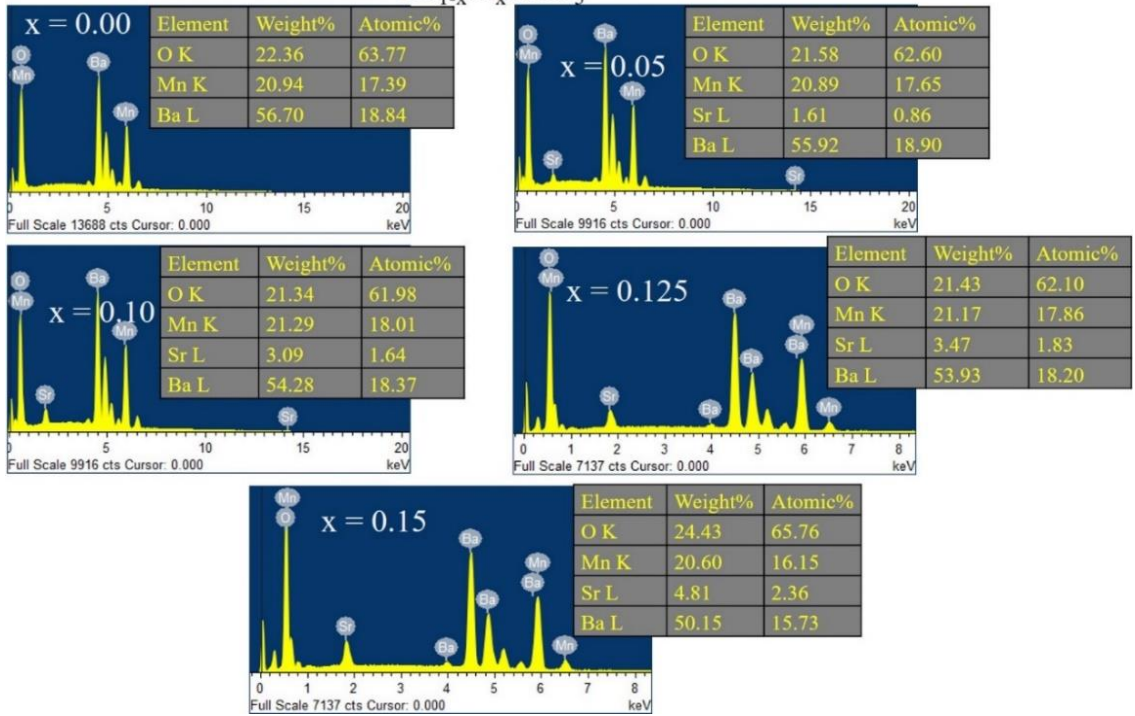

Figure S4. EDAX spectra of Ba$_{1-x}$Sr$_x$MnO$_3$: x = 0.00 to 0.15

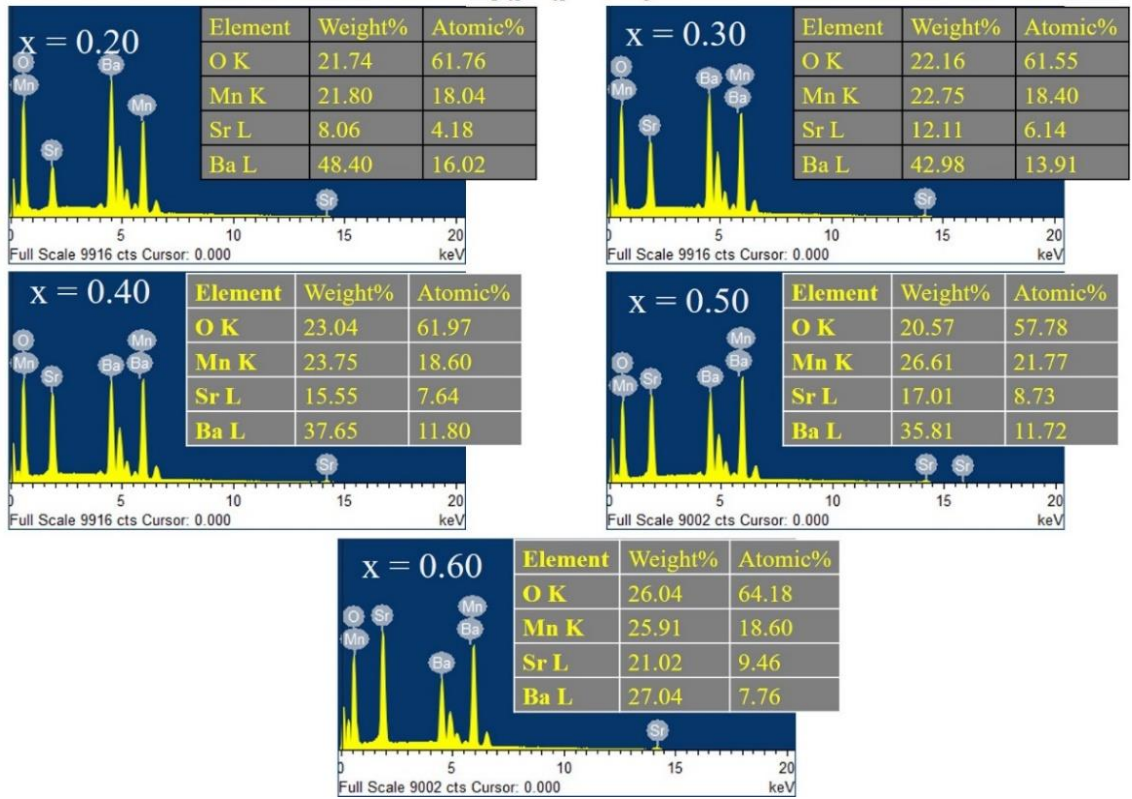

Figure S5. EDAX spectra of Ba$_{1-x}$Sr$_x$MnO$_3$: x = 0.20 to 0.60

Table S2: EDAX data for $Ba_{1-x}Sr_xMnO_3$: x = 0.00 to 0.60

| $Ba_{1-x}Sr_xMnO_3$ | | | | | |
|---|---|---|---|---|---|
| x (Planned) | 0.00 | 0.05 | 0.10 | 0.125 | 0.15 |
| Element | Atomic% | Atomic% | Atomic% | Atomic% | Atomic% |
| O K | 63.77 | 62.60 | 61.98 | 62.10 | 65.76 |
| Mn K | 17.39 | 17.65 | 18.01 | 17.86 | 16.15 |
| Sr L | 0.00 | 0.86 | 1.64 | 1.83 | 2.36 |
| Ba L | 18.84 | 18.90 | 18.37 | 18.20 | 15.73 |
| x (Obtained from EDAX) | 0.00 | 0.044 | 0.082 | 0.091 | 0.130 |
| | | | | | |
| x (Planned) | 0.20 | 0.30 | 0.40 | 0.50 | 0.60 |
| Element | Atomic% | Atomic% | Atomic% | Atomic% | Atomic% |
| O K | 61.76 | 61.55 | 61.97 | 57.78 | 64.18 |
| Mn K | 18.04 | 18.40 | 18.60 | 21.77 | 18.60 |
| Sr L | 4.18 | 6.14 | 7.64 | 8.73 | 9.46 |
| Ba L | 16.02 | 13.91 | 11.80 | 11.72 | 7.76 |
| x (Obtained from EDAX) | 0.207 | 0.306 | 0.393 | 0.427 | 0.549 |

**Magnetism**

Figure S6 shows magnetization curves measured down to 2 K. As the temperature is lowered below $T_N$, ZFC and FC curves in M(T) show significant bifurcation, especially below $T^* \sim 43$ K exhibiting a peak-like feature in ZFC curve. These signatures suggest that $Ba_{1-x}Sr_xMnO_3$ possibly undergoes a phase transition from antiferromagnetic (AFM) state to ferrimagnetic or canted AFM [Phys. Rev. B 102, 134436 (2020)], arising from the multi valency of Mn. Note that the magnetic phase transition for 2H phase compositions (x = 0 and 0.05) which is expected at 59 K might be overshadowed by $T^*$ [Chem. Mater. 12, 831 (2000)].

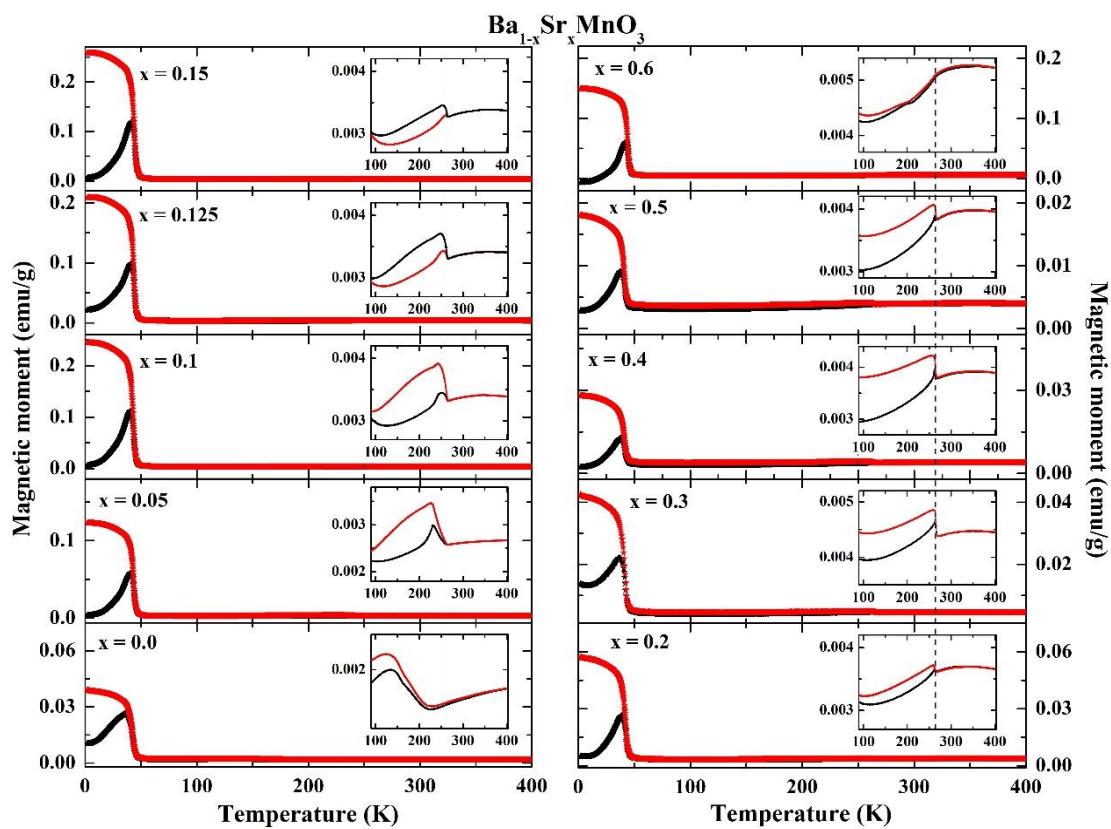

Figure S6. Magnetization as a function of temperature for $Ba_{1-x}Sr_xMnO_3$: x = 0.00 to 0.60.

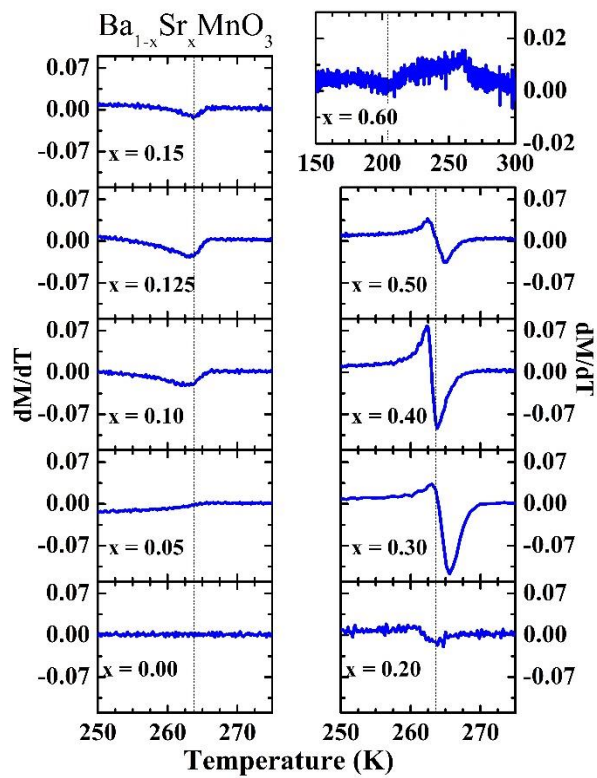

Figure S7. Temperature derivative of magnetization for $Ba_{1-x}Sr_xMnO_3$: x = 0.00 to 0.60.

**Raman spectroscopy**

Table S3. Symmetry assignment of the Raman active phonon modes and their frequencies in cm$^{-1}$ unit) in Ba$_{1-x}$Sr$_x$MnO$_3$: x = 0.00 & 0.05 (2H phase).

| 2H: *P6₃mc*: $\Gamma'_{2H} = 4A_1 + 5E_1 + 5E_2$ |||||
|---|---|---|---|---|
| 9R: *R-3m*: $\Gamma'_{9R} = 4A_{1g} + 5E_g$ |||||
| 4H: *P6₃/mmc*: $\Gamma'_{4H} = 2A_{1g} + 2E_{1g} + 4E_{2g}$ |||||
| Weak modes $\omega$ (cm$^{-1}$) |||| Symmetry |
| BaMnO$_3$ || Ba$_{0.95}$Sr$_{0.05}$MnO$_3$ || |
| 80 K | 300 K | 80 K | 300 K | |
| 193 | -- | 195 | -- | -- |
| 197 | 191 | 200 | 192 | -- |
| 221 | 220 | 222 | 219 | -- |
| 272 | -- | 271 | -- | -- |
| 279 | 275 | 280 | 275 | -- |
| 282 | -- | -- | -- | -- |
| 343 | 344 | 344 | 342 | $E_2$ |
| 366 | -- | -- | -- | $A_1$ |
| Second order modes |||||
| 557 | 551 | 560 | 554 | -- |
| 567 | 561 | 569 | 562 | -- |
| 575 | Absent | -- | -- | -- |
| 580 | 573 | 582 | 576 | -- |
| 590 | 586 | 590 | 587 | -- |

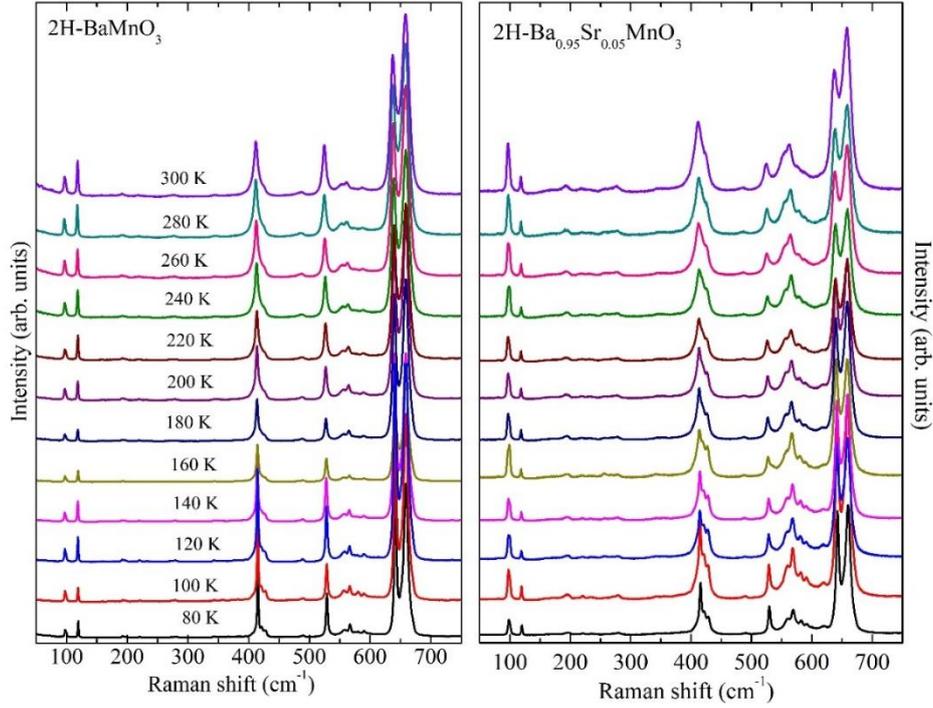

Figure S8. Raman spectra collected at a few temperatures for $Ba_{1-x}Sr_xMnO_3$: x = 0.00 & 0.05.

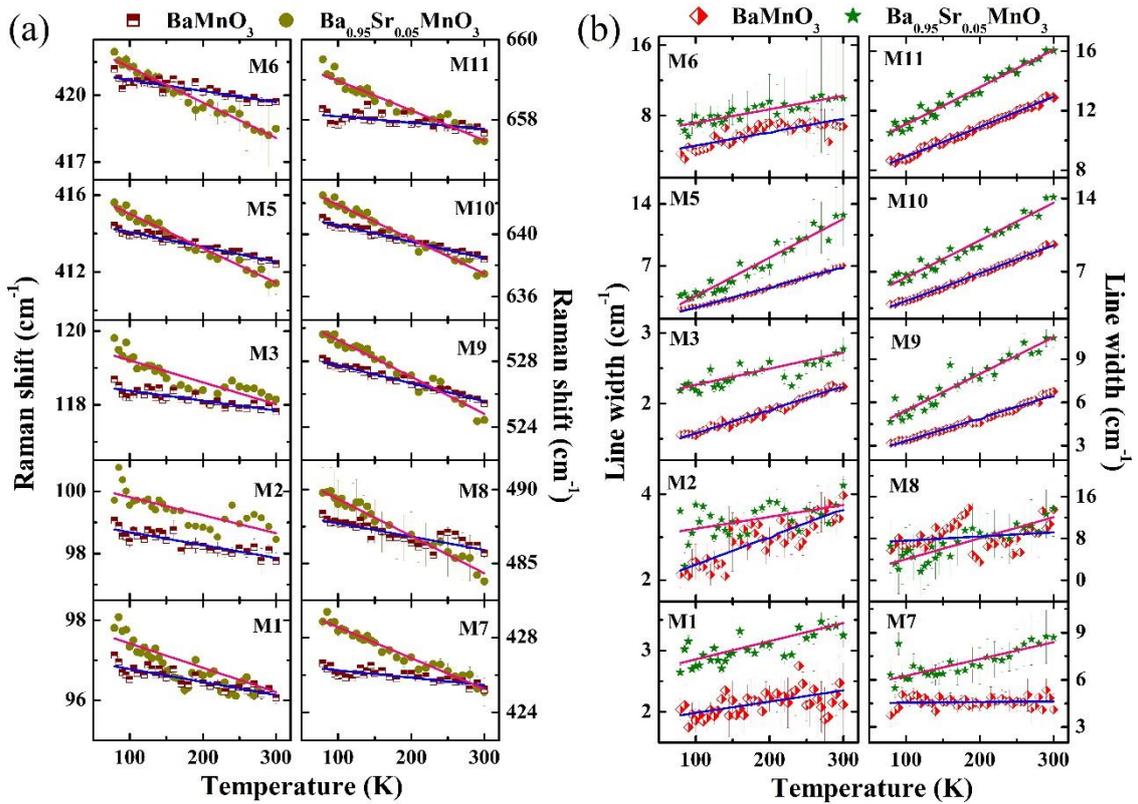

Figure S9. Frequency (a) and linewidth (b) of phonons as a function of temperature for $Ba_{1-x}Sr_xMnO_3$: x = 0.00 & 0.05. Solid lines in (a) and (b) are anharmonic fittings with Eqns. 2 and 3, respectively, explained in main text.

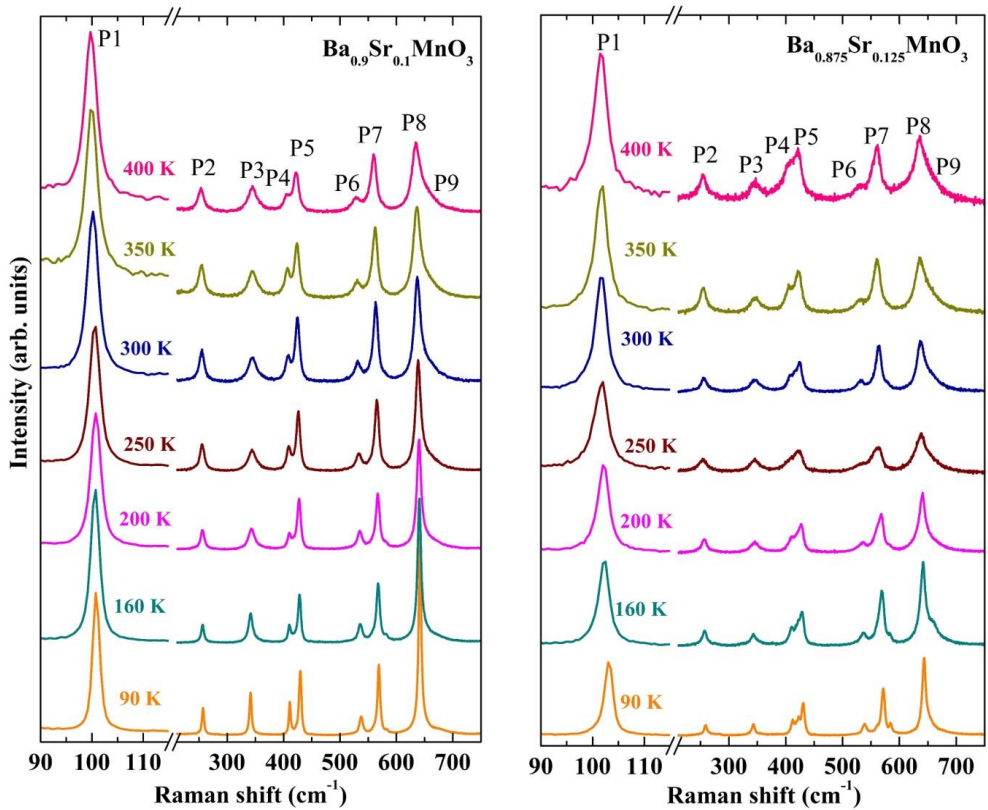

Figure S10. Raman spectra collected at a few temperatures for $Ba_{1-x}Sr_xMnO_3$: x = 0.10 & 0.125.

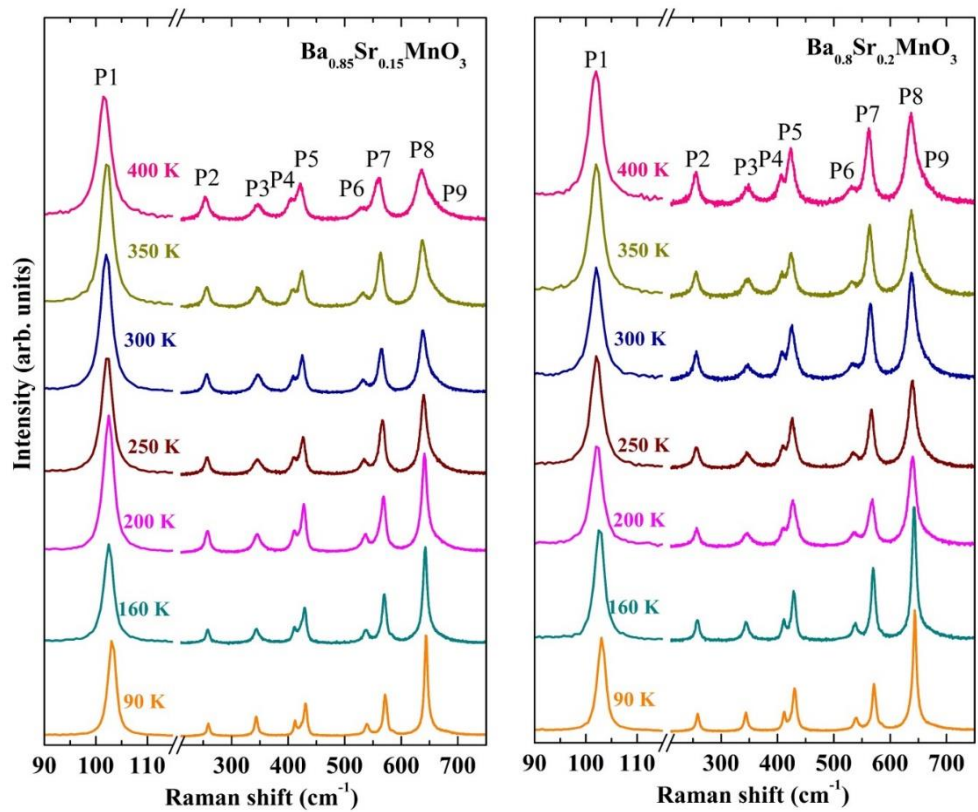

Figure S11. Raman spectra collected at a few temperatures for $Ba_{1-x}Sr_xMnO_3$: x = 0.15 & 0.20.

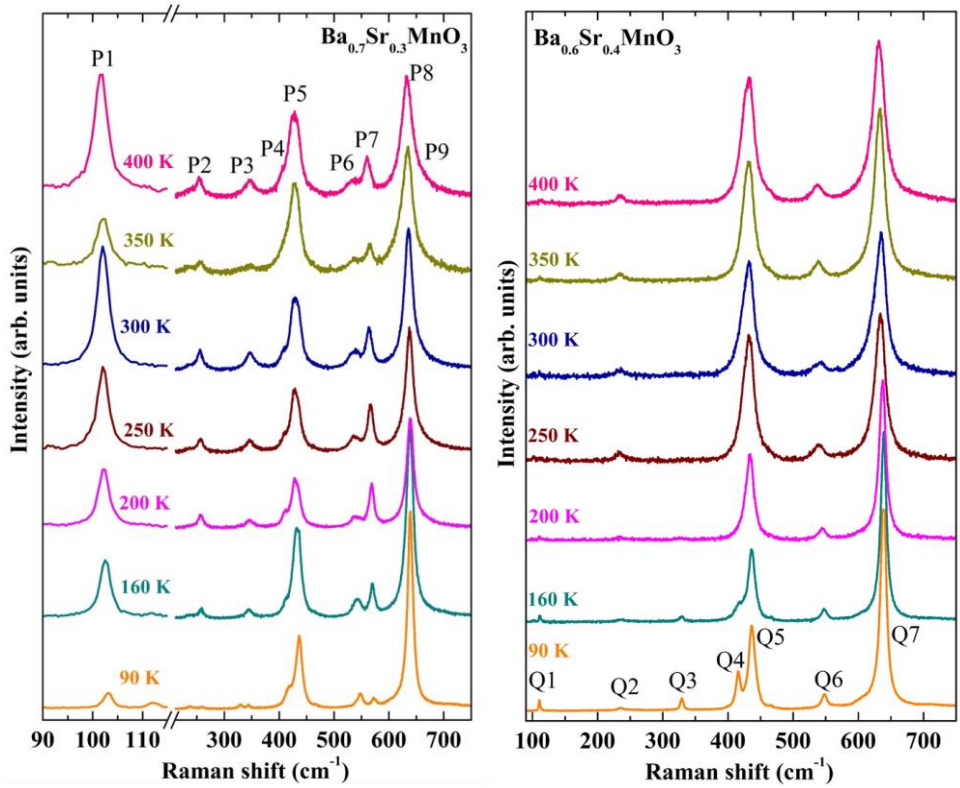

Figure S12. Raman spectra collected at a few temperatures for $Ba_{1-x}Sr_xMnO_3$: x = 0.30 & 0.40.

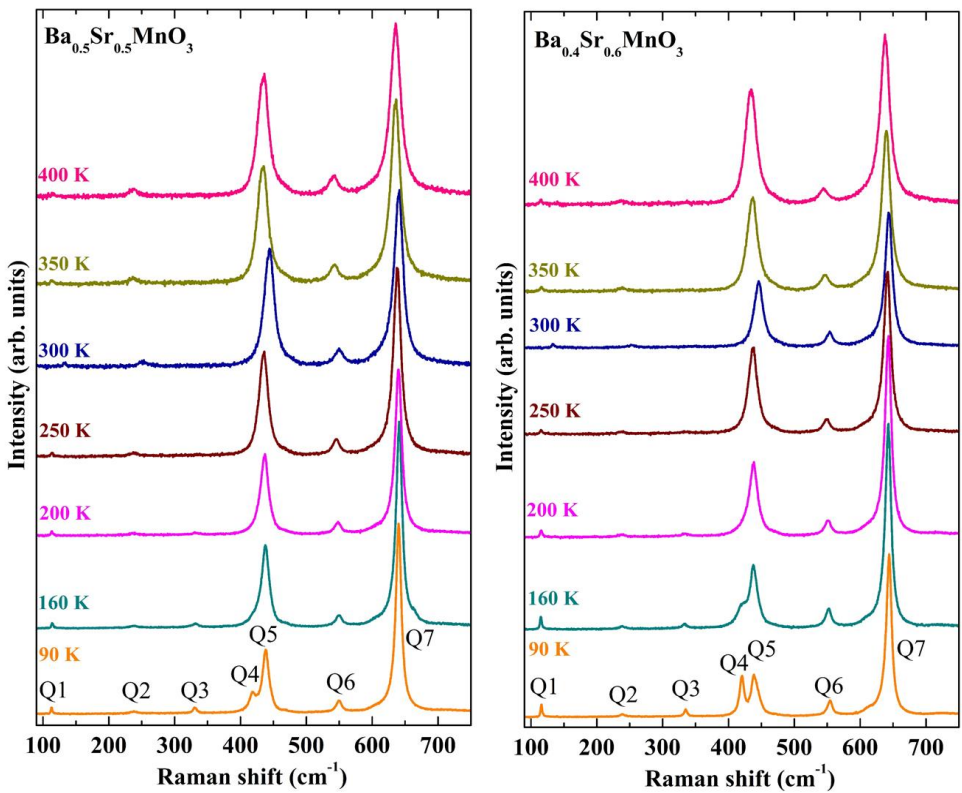

Figure S13. Raman spectra collected at a few temperatures for $Ba_{1-x}Sr_xMnO_3$: x = 0.50 & 0.60.

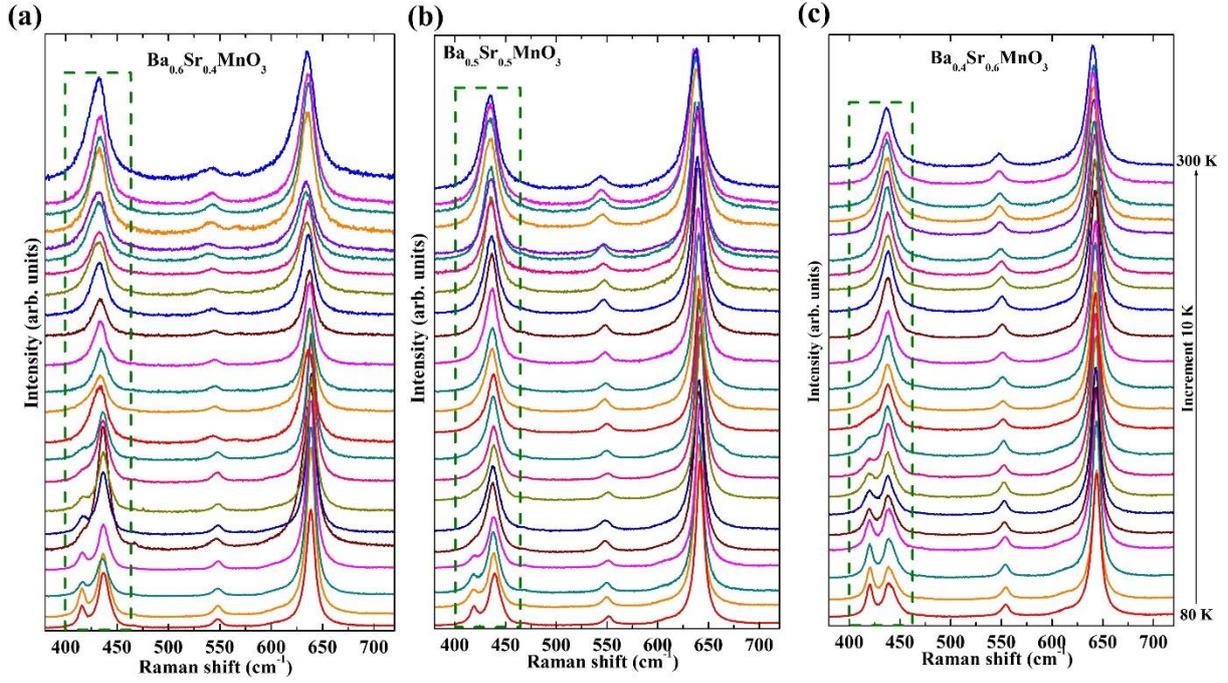

Figure S14. Raman spectra of 4H compositions (x = 0.40-0.60) are plotted in the high-frequency range to capture the temperature dependence of Q4 (~ 419 cm$^{-1}$) and Q5 (~ 432 cm$^{-1}$) modes clearly, as shown by the dashed boxes.

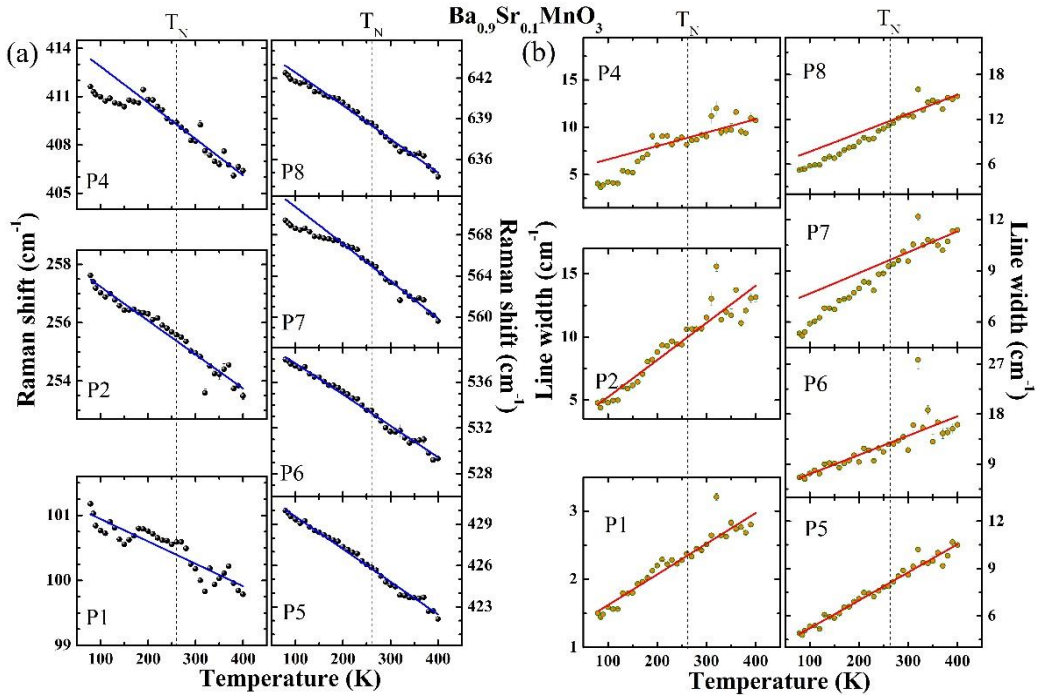

Figure S15. Frequency (a) and linewidth (b) of phonons as a function of temperature for $Ba_{1-x}Sr_xMnO_3$: x = 0.10. Solid lines in (a) and (b) are anharmonic fittings with Eqns. 2 and 3, respectively, explained in main text. A clear deviation from the anharmonic trend is seen for the modes P4, P7, and P8.

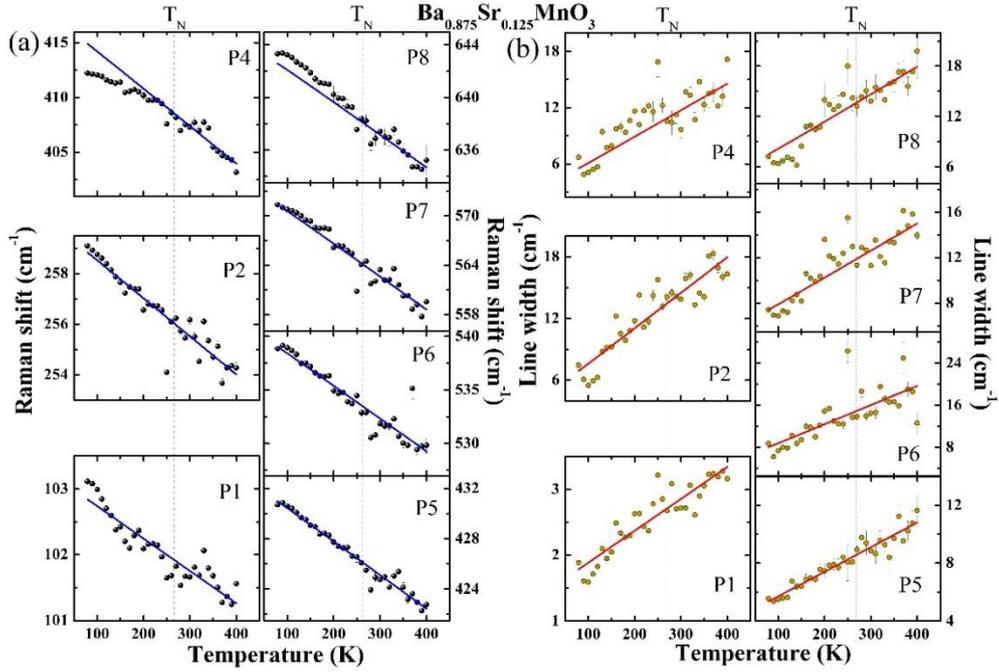

Figure S16. Frequency (a) and linewidth (b) of phonons as a function of temperature for $Ba_{1-x}Sr_xMnO_3$: $x = 0.125$. Solid lines in (a) and (b) are anharmonic fittings with Eqns. 2 and 3, respectively, explained in main text. A deviation from the anharmonic trend is seen for the modes P4, P7, and P8.

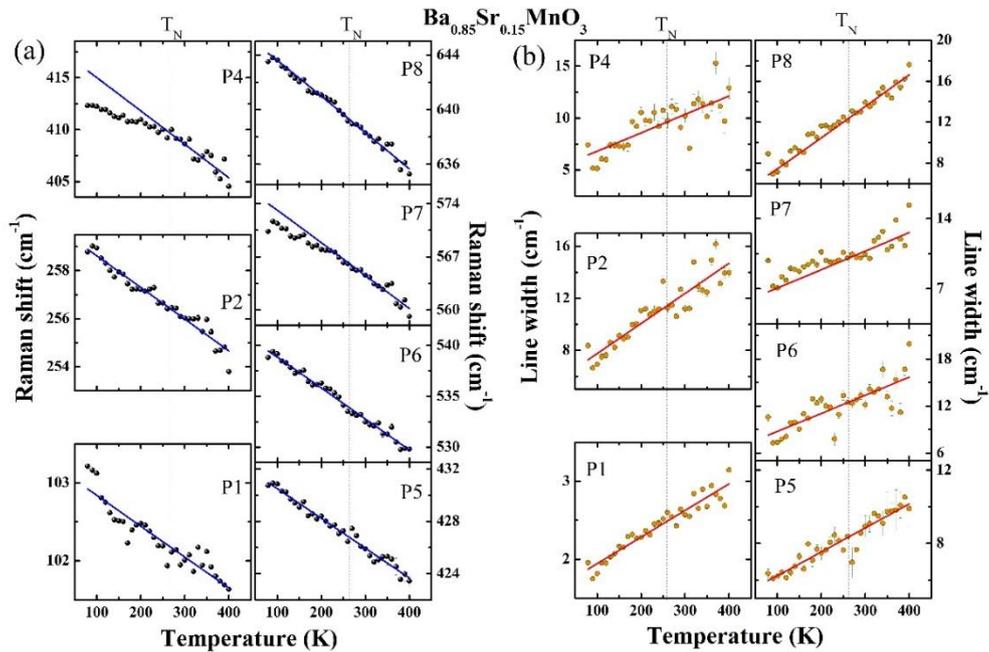

Figure S17. Frequency (a) and linewidth (b) of phonons as a function of temperature for $Ba_{1-x}Sr_xMnO_3$: $x = 0.15$. Solid lines in (a) and (b) are anharmonic fittings with Eqns. 2 and 3, respectively, explained in main text. A deviation from the anharmonic trend is seen for the modes P4, P7, and P8.

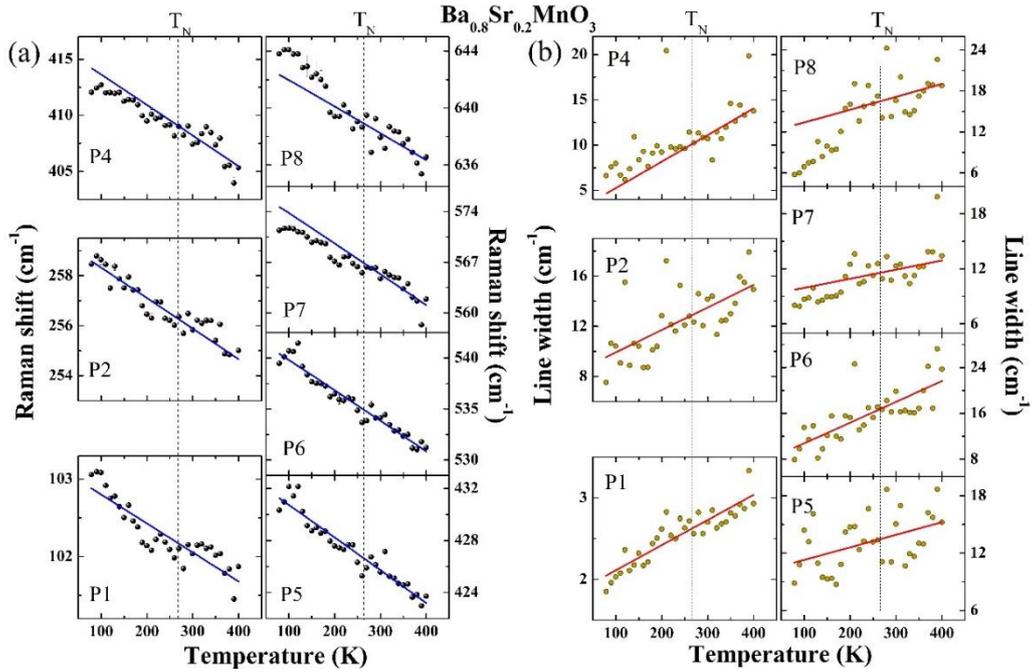

Figure S18. Frequency (a) and linewidth (b) of phonons as a function of temperature for Ba$_{1-x}$Sr$_x$MnO$_3$: x = 0.20. Solid lines in (a) and (b) are anharmonic fittings with Eqns. 2 and 3, respectively, explained in main text. A deviation from the anharmonic trend is seen for the modes P4, P7, and P8.

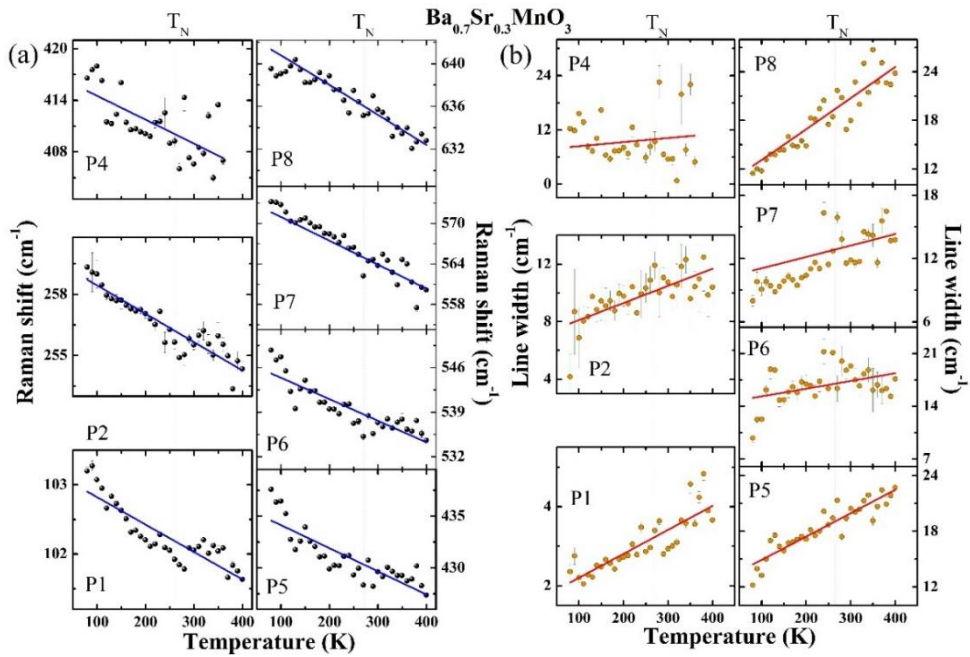

Figure S19. Frequency (a) and linewidth (b) of phonons as a function of temperature for Ba$_{1-x}$Sr$_x$MnO$_3$: x = 0.30. Solid lines in (a) and (b) are anharmonic fittings with Eqns. 2 and 3, respectively, explained in main text. A deviation from the anharmonic trend is seen for the modes P4, P7, and P8.

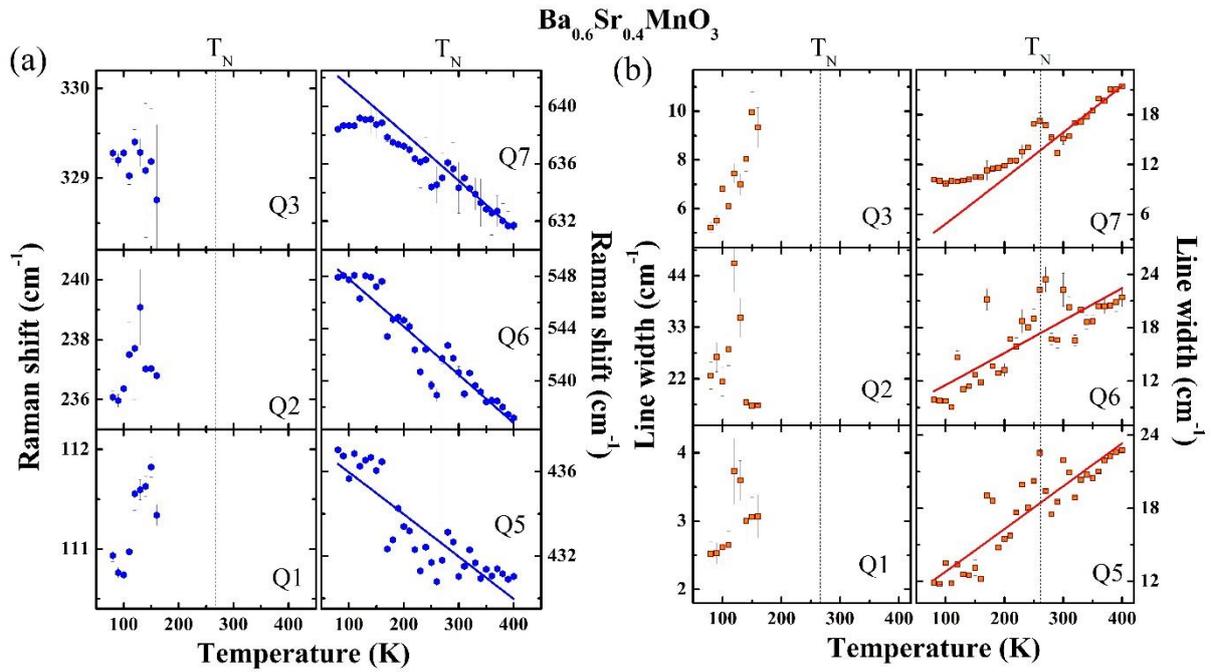

Figure S20. Frequency (a) and linewidth (b) of phonons as a function of temperature for Ba$_{1-x}$Sr$_x$MnO$_3$: x = 0.40. Solid lines in (a) and (b) are anharmonic fittings with Eqns. 2 and 3, respectively, explained in main text. Mode Q7 deviates from the expected anharmonic trend.

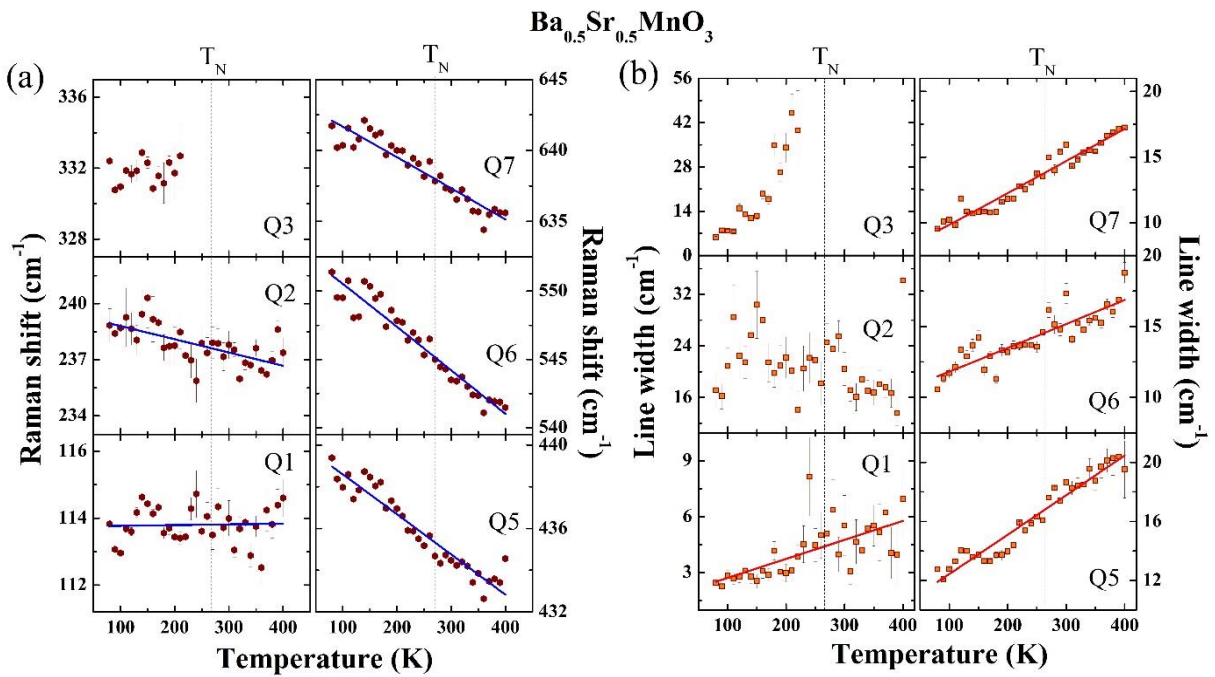

Figure S21. Frequency (a) and linewidth (b) of phonons as a function of temperature for Ba$_{1-x}$Sr$_x$MnO$_3$: x = 0.50. Solid lines in (a) and (b) are anharmonic fittings with Eqns. 2 and 3, respectively, explained in main text.

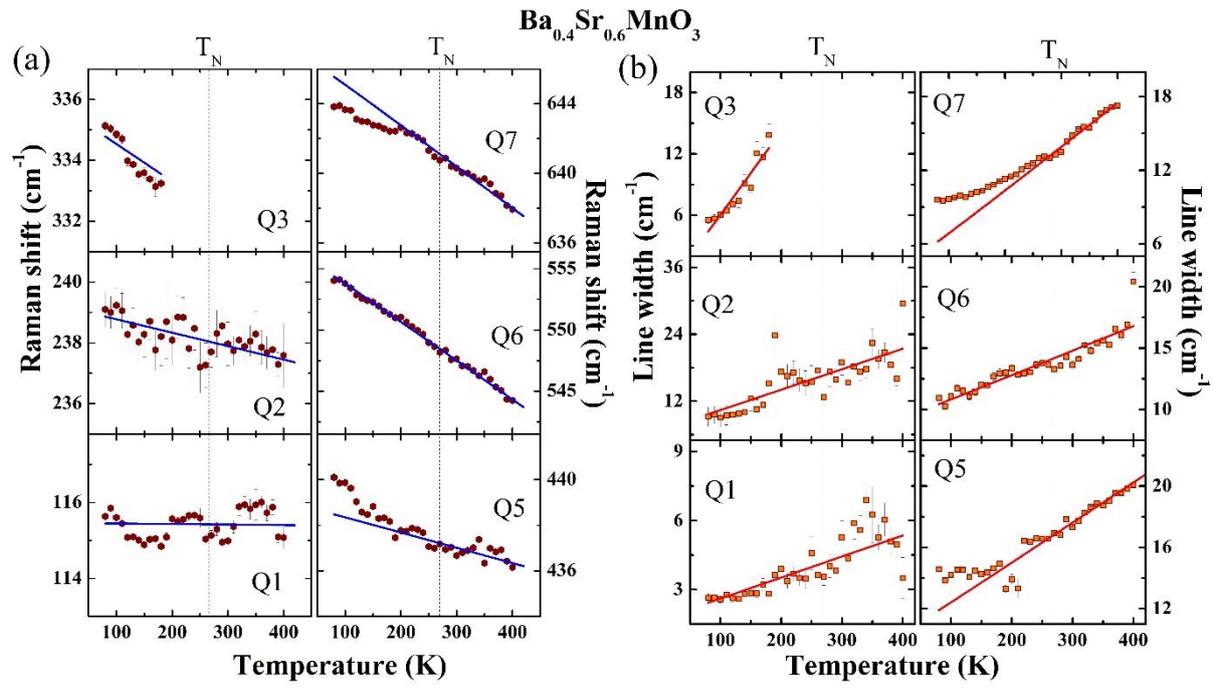

Figure S22. Frequency (a) and linewidth (b) of phonons as a function of temperature for $Ba_{1-x}Sr_xMnO_3$: x = 0.60. Solid lines in (a) and (b) are anharmonic fittings with Eqns. 2 and 3, respectively, explained in main text. Mode Q7 deviates from the expected anharmonic trend.

**Thermal expansion-temperature-dependent XRD:**

2H phase compositions of $Ba_{1-x}Sr_xMnO_3$ (x = 0 and 0.05) exhibit a structural phase transition with temperature at $T_C \sim 130$ K as shown in Fig. S23 (b, d). The crystal symmetry changes from $P6_3mc$ at high-temperature to $P6_3cm$ at low temperature [Phys. Rev. B 92, 134308 (2015)]. Solid lines in Figs. S24-S26 (b, d) are fittings with thermal expansion equation which can be written as:

$$a(T) = a_0\left[1 + \frac{be^{\frac{d}{T}}}{T(e^{\frac{d}{T}}-1)^2}\right] \quad \text{and} \quad c(T) = c_0\left[1 + \frac{fe^{\frac{g}{T}}}{T(e^{\frac{g}{T}}-1)^2}\right] \quad (S2)$$

where, $a_0$ and $c_0$ are the in-plane and out-of-plane lattice constants at 0 K, whereas b, d, f, and g are fitting parameters [Charles Kittel, "Introduction to Solid State Physics", seventh ed., Wiley, New York, 2003].

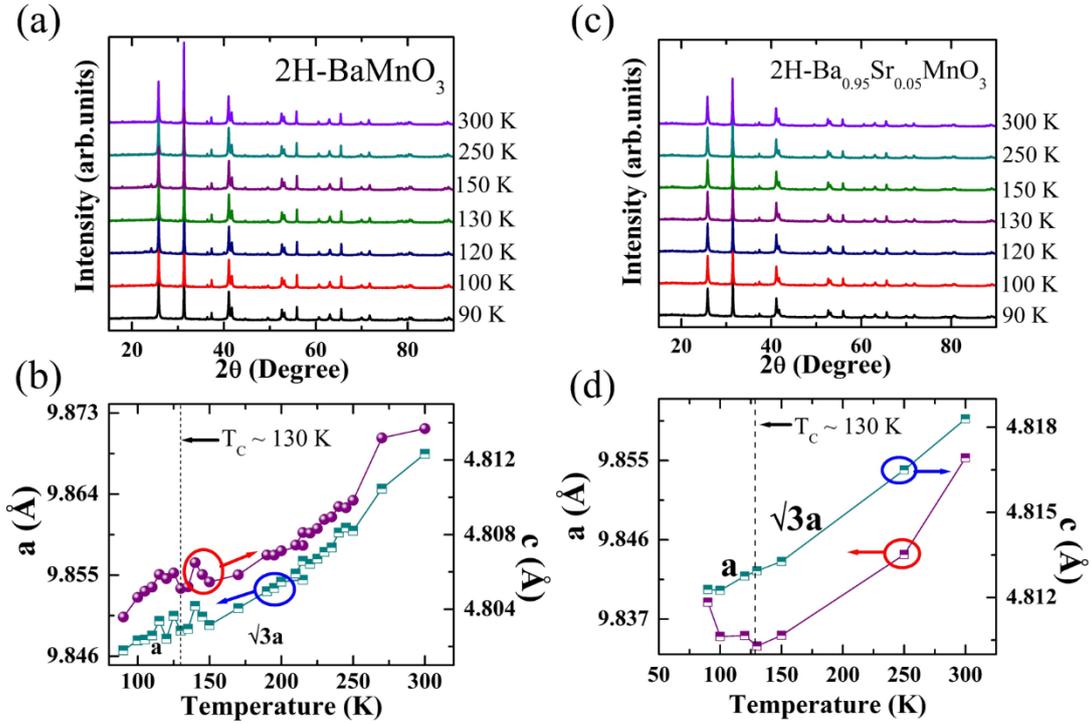

Figure S23: (a, c) X-ray diffraction patterns at a few temperatures and (b, d) lattice parameters as a function of temperature for $Ba_{1-x}Sr_xMnO_3$: x = 0.00 & 0.05 in its 2H phase. The vertical dashed line in (b) and (d) represents the structural phase transition temperature.

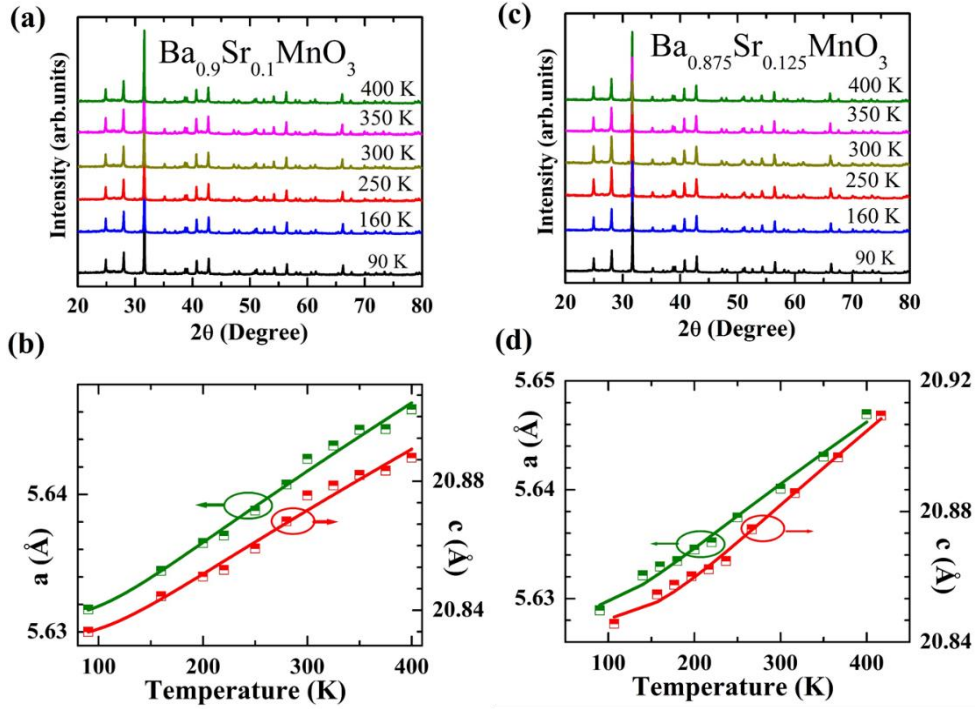

Figure S24: (a, c) X-ray diffraction patterns at a few temperatures and (b, d) lattice parameters as a function of temperature for $Ba_{1-x}Sr_xMnO_3$: x = 0.10 & 0.125 in its 9R phase.

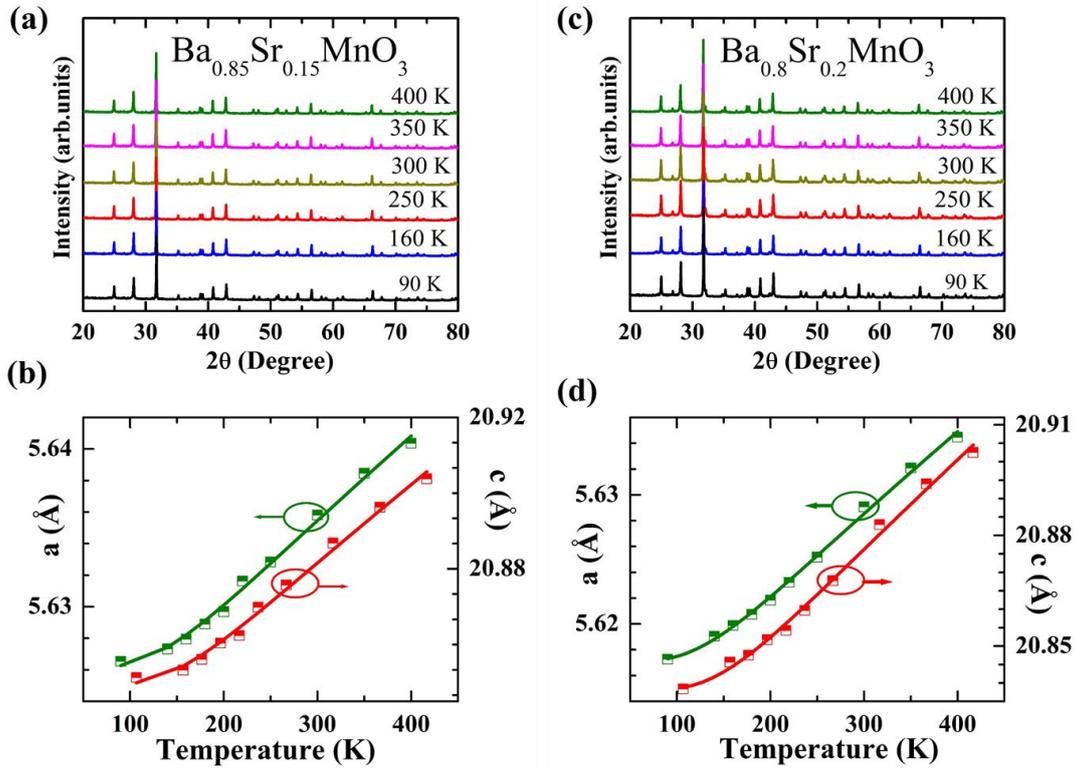

Figure S25: (a, c) X-ray diffraction patterns at a few temperatures and (b, d) lattice parameters as a function of temperature for $Ba_{1-x}Sr_xMnO_3$: x = 0.15 & 0.20 in its 9R phase.
18

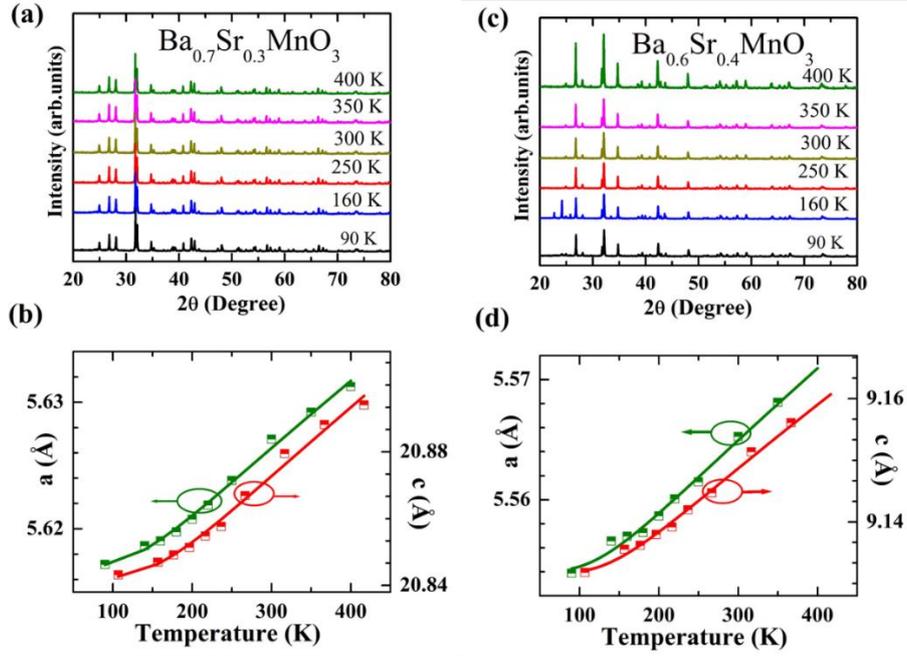

Figure S26: (a, c) X-ray diffraction patterns at a few temperatures and (b, d) lattice parameters as a function of temperature for $Ba_{1-x}Sr_xMnO_3$: x = 0.30 & 0.40 its 9R and 4H phase, respectively.

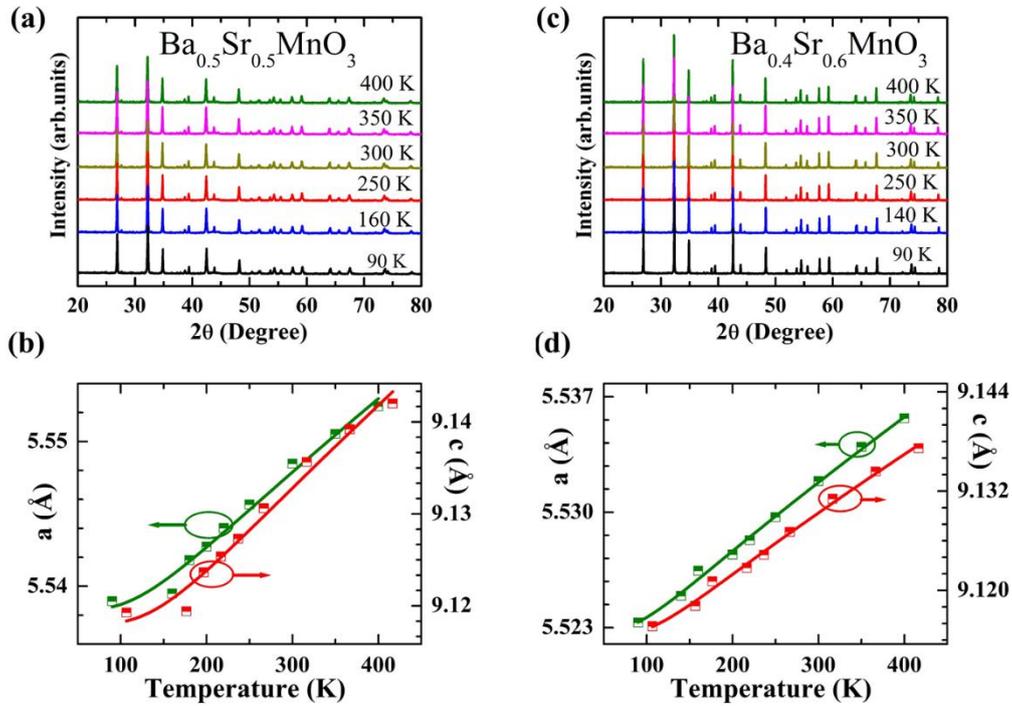

Figure S27: (a, c) X-ray diffraction patterns at a few temperatures and (b, d) lattice parameters as a function of temperature for $Ba_{1-x}Sr_xMnO_3$: x = 0.50 & 0.60 in its 4H phase.

**Spin-phonon coupling of P4, P7, P8, and Q7 modes:**

The modes P4, P7, P8, and Q7 show clear deviation from their expected anharmonic trend below $T_N$ indicating the presence of spin-phonon coupling. We have estimated the strength of spin-phonon coupling using the relation (as discussed in the main text):

$$\lambda = - \frac{\omega(T_{Low}) - \omega_{anh}(T_{Low})}{[\Phi(T_{Low}) - \Phi(2T_N)]S^2} \quad (S3)$$

where $\omega(T_{Low})$ is the experimental phonon frequency at the lowest temperature recorded ($T_{Low} \sim 80$ K in our case) while the $\omega_{anh}(T_{Low})$ is the corresponding anharmonic estimate of the phonon frequency at the same temperature. $\Phi(T)$ is the short-range order parameter and $S$ is the spin. The method is described in the main text (Eqns. 4 and 5 in the main text). We observed a significant variation in λ with varying Sr content across the various crystallographic phases.

Table S4. Spin-phonon coupling constant (λ) for P4, P7, and P8 modes in the 9R phase (x = 0.10 to 0.30) and Q7 in 4H phase (x = 0.40 to 0.60) of $Ba_{1-x}Sr_xMnO_3$.

| Composition (x) ➔ | 0.10 | 0.125 | 0.15 | 0.20 | 0.30 |
|---|---|---|---|---|---|
| Mode | λ (cm$^{-1}$) | | | | |
| P4 | 0.9 | 1.4 | 1.7 | 1.1 | 0.8 |
| P7 | 1.1 | 0.0 | 1.9 | 1.7 | 0.8 |
| P8 | 0.3 | 0.0 | 0.3 | 0.8 | 0.9 |
| Composition (x) ➔ | 0.40 | 0.50 | 0.60 | | |
| Mode | λ (cm$^{-1}$) | | | | |
| Q7 | 1.9 | 0.0 | 0.9 | | |